\documentclass{aa} 
\usepackage{graphicx}
\usepackage{txfonts}
\usepackage[usenames]{color} 
\usepackage{tabularx} 
\usepackage{footnote}
\makesavenoteenv{table*}
\usepackage{rotating}
\usepackage{float}
\usepackage{longtable} 
\usepackage{lscape} 

\usepackage{array}
\newcolumntype{H}{>{\setbox0=\hbox\bgroup}(c)<{\egroup}@{}}
\usepackage{soul}
\usepackage{scalerel}
\usepackage{tikz}
\usetikzlibrary{svg.path}
\definecolor{orcidlogocol}{HTML}{A6CE39}
\tikzset{
  orcidlogo/.pic={
    \fill[orcidlogocol] svg{M256,128c0,70.7-57.3,128-128,128C57.3,256,0,198.7,0,128C0,57.3,57.3,0,128,0C198.7,0,256,57.3,256,128z};
    \fill[white] svg{M86.3,186.2H70.9V79.1h15.4v48.4V186.2z}
                 svg{M108.9,79.1h41.6c39.6,0,57,28.3,57,53.6c0,27.5-21.5,53.6-56.8,53.6h-41.8V79.1z M124.3,172.4h24.5c34.9,0,42.9-26.5,42.9-39.7c0-21.5-13.7-39.7-43.7-39.7h-23.7V172.4z}
                 svg{M88.7,56.8c0,5.5-4.5,10.1-10.1,10.1c-5.6,0-10.1-4.6-10.1-10.1c0-5.6,4.5-10.1,10.1-10.1C84.2,46.7,88.7,51.3,88.7,56.8z};
  }
}

\newcommand\orcid[1]{\href{https://orcid.org/#1}{\mbox{\scalerel*{
\begin{tikzpicture}[yscale=-1,transform shape]
\pic{orcidlogo};
\end{tikzpicture}
}{|}}} \href{#1}{#1}}
\usepackage[breaklinks=true,colorlinks, linkcolor=blue,urlcolor=blue,citecolor=blue]{hyperref}

\usepackage{threeparttablex} 

\usepackage[switch]{lineno}

\begin{document} 

   \title{Comparing observations of the closely located JUICE and STEREO-A spacecraft during the widespread solar energetic particle event of 2024 May 13}

   \titlerunning{JUICE and STEREO-A observations of the widespread SEP event on 2024 May 13}


   \author{L.~Rodríguez-García\inst{1,2} 
          \and 
          E.~Palmerio\inst{3}
          \and
          M.~Pinto\inst{4}
          \and 
          N.~Dresing\inst{5}
          \and 
          C.~M.~S.~Cohen\inst{6} 
          \and 
          R.~Gómez-Herrero\inst{2} 
          \and 
          J.~Gieseler\inst{5}
          \and 
          F.~Santos\inst{4}
          \and 
          F.~Espinosa Lara\inst{2}
          \and
          I. Cernuda\inst{2}
          \and
          M.~Mewes\inst{1,7}
          \and
          C.~Vallat\inst{1}
          \and
          O.~Witasse\inst{8} 
          \and
          N.~Altobelli\inst{1}
          }
          
   \institute{
        European Space Agency (ESA), European Space Astronomy Centre (ESAC), Camino Bajo del Castillo s/n, 28692 Villanueva de la Cañada, Madrid, Spain \\
        \email{laura.rodriguezgarcia@esa.int}
        \and Universidad de Alcalá, Space Research Group (SRG-UAH), Plaza de San Diego s/n, 28801 Alcalá de Henares, Madrid, Spain
        \and 
        Predictive Science Inc., San Diego, CA 92121, USA
        \and
        Laboratory for Instrumentation and Experimental Particle Physics (LIP), 1649-003 Lisboa, Portugal
        \and
        Department of Physics and Astronomy, University of Turku, FI-20014 Turku, Finland 
        \and
        California Institute of Technology, Pasadena, CA 91125, USA
        \and
        Institut f{\"u}r Experimentelle und Angewandte Physik, University of Kiel, 24118 Kiel, Germany
        \and
        European Space Research and Technology Centre, European Space Agency, 2201 AZ Noordwijk, The Netherlands
        }

   \date{Received xxx, 2025; accepted xxxxx, 2025}


 
  \abstract
  {JUICE was launched in April 2023, and it is now in its cruise phase to Jupiter, where it is scheduled to arrive in July 2031. JUICE carries a radiation monitor, namely the RADiation hard Electron Monitor (RADEM) to measure protons, electrons, and ions, detecting particles coming mainly from the anti-Sun direction. On 2024 May 13, a large solar energetic particle (SEP) event took place in association with an eruption close to the western limb of the Sun as seen from Earth. Providentially, at that time JUICE was located very close to STEREO-A, being separated by 0.13~au in radial distance, 0.3$^{\circ}$ in latitude, and 1.6$^{\circ}$ in longitude.   }
   {Our main aims are to characterise the observations within the interplanetary (IP) context through which SEPs propagated to near-Earth, JUICE, and STEREO-A observers and to perform a first comparison of the energetic particle instruments on board the JUICE and STEREO-A spacecraft.  } 
   {We analysed the IP context using in-situ measurements and studied the proton anisotropies measured by near-Earth spacecraft and STEREO-A. We focused on an isotropic period during the decay phase of the SEP event to compute the proton energy spectrum. We fit the STEREO-A spectrum and compared it to that measured by SOHO and JUICE. 
}
   {The proton spectral indices measured by JUICE, SOHO, and STEREO-A were found to be similar. The proton fluxes measured by RADEM agree with those from STEREO-A, with a deviation of less than 25\%.  }
   {The RADEM instrument aboard JUICE is a valuable tool for measuring SEP events in the heliosphere, providing an excellent opportunity to study and characterise the energetic particle environment in the solar wind between 0.65 and 5.2~au. The intercalibration factors between the fluxes measured by STEREO-A and JUICE at the effective energies of 6.9 MeV, 13.3 MeV, 21.6 MeV, and 31.2 MeV are 1.02, 1.23, 1.12, and 0.95, respectively. These intercalibration factors are valid only until 2024 July 10, when the configuration of the RADEM instrument was changed.}
   
   \maketitle


\section{Introduction}
\label{sec:Introduc}

The Sun and its sphere of influence---the heliosphere---are characterised by a variable particle environment shaped by the dynamic solar activity. Of particular interest are so-called solar energetic particle (SEP) events, periods during which a certain region of the heliosphere is affected by enhanced fluxes of energetic protons, electrons, and heavy ions, with the potential of damaging spacecraft electronics and delivering increased radiation to astronauts in orbit \citep[e.g.][]{Jiggens2014}. SEPs are mainly accelerated in association with solar eruptions such as flares and coronal mass ejections (CMEs), and are spread outwards from the Sun via a number of mechanisms that may involve particle transport along or perpendicular to magnetic field lines \citep[e.g.][]{Dresing2014}. To enhance the physical understanding and predictive capabilities of SEPs \citep[the current status of SEP modelling has been recently reviewed by][]{Whitman2023}, the heliophysics research community has concentrated its efforts on studying events detected by multiple spacecraft positioned at widely separated locations in the heliosphere, which can provide additional insights on particle acceleration and transport at a ``global'' level \citep[e.g.][]{Kollhoff2021, Rodriguez-Garcia2021, Lario2022}. Given the enormous spatial scales involved and the exiguous number of probes covering different regions of the solar system, many analyses of SEP measurements in the heliosphere have been possible thanks to data from planetary missions, which have been employed either in statistical studies \citep[e.g.][]{Rodriguez-Garcia2023a, Sanchez-Cano2023} or to take advantage of multi-spacecraft observations of a single event \citep[e.g.][]{Palmerio2021, Dresing2023, Khoo2024,Dresing2025}.

\begin{figure*}[th!]
\centering
  \includegraphics[width=0.9\linewidth]{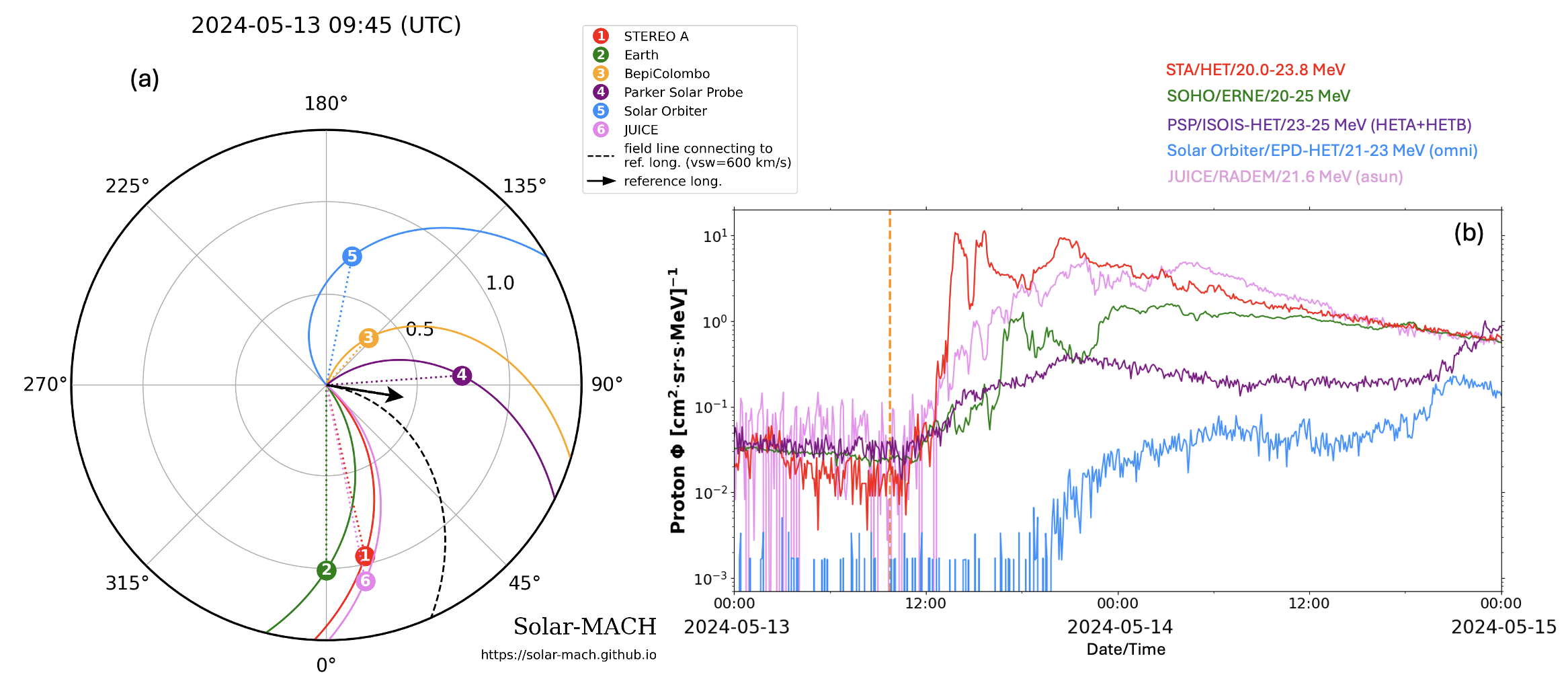}
     \caption{Spatial distribution of spacecraft and their magnetic connectivity at 09:45 UT on May 13, 2024, along with SEP observations from multiple spacecraft. (a) The spacecraft constellation was produced using the Solar-MACH tool \citep{Gieseler2023}, which is accessible online at \url{https://doi.org/10.5281/zenodo.7016783}. The solar wind speeds applied at various positions are derived from a combination of in-situ measurements and estimated values, as detailed in the main text. (b) Proton intensities near 22 MeV recorded by the different spacecraft are shown. The orange vertical line marks the time of the flare’s soft X-ray peak (around 09:44 UT), which is associated with the observed SEP event.}
     \label{fig:solar_mach}
\end{figure*}

\begin{table*}
\centering
\caption{List of linearly combined RADEM/PDH channels with their respective proton energy ranges and effective energy.}
\begin{tabular}{cccccc}
\hline
Channel Combination             & Energy Range (MeV)  & 
Effective Energy (MeV) & GdE (cm$^2\cdot$sr$\cdot$MeV) & $\delta^-_G$ (\%) & $\delta^+_G$ (\%)\\ \hline
\hline
(1)&(2)&(3)&(4)&(5)&(6)\\
\hline
{1x[Proton\_Bin\_1] - 1x[Proton\_Bin\_3]} & 5.35--14.4 & 6.9 & 0.214 & -1.22 & 3.00     \\
{1x[Proton\_Bin\_2] - 1x[Proton\_Bin\_4]} & 8.75--22.8 & 13.3 & 0.837 & -9.31 & 24.6    \\
{1x[Proton\_Bin\_3] - 1x[Proton\_Bin\_5]} & 14.5--37.4 & 21.6 & 1.22 & -6.53 & 17.93   \\
{1x[Proton\_Bin\_4] - 1x[Proton\_Bin\_5]} & 22.8--36.6 & 31.2 & 0.844 & -2.91 & 8.58 \\\hline

\label{table:RADEMchannels}
\end{tabular}
\footnotesize{\textbf{Notes.} Col. 1: Linear combination of proton channels used to create differential channels. Col. 2: Energy range of the proton channel.  Col. 3: Effective energy obtained with the bow-tie method. Col. 4: Mean value of the geometric factor distribution calculated with the bow-tie method. Cols. 5 and 6: 5th and 95th percentile of the geometric factor distribution subtracted from the mean value in percentage calculated with the bow-tie method. Details given in the main text. }
\end{table*}
The JUpiter ICy moons Explorer \citep[JUICE;][]{Grasset2013} spacecraft was launched on 2023 April 14 towards the largest planet in the solar system, to perform detailed studies of its environment and that of its three ocean-bearing moons---Europa, Ganymede, and Callisto. It is equipped with remote-sensing, geophysical, and in-situ instruments, and is currently on its way towards the Jovian system with expected orbit insertion in July 2031. Amongst its suite of instrumentation, JUICE carries the RADiation hard Electron Monitor \citep[RADEM;][]{Pinto2020, RADEM2025}, which is able to measure protons, electrons, and heavier ions to characterise the high-radiation Jovian particle environment. Apart from its planned planetary investigations, RADEM is operational over the whole mission's 8-year cruise phase, thus providing an excellent opportunity to study and characterise the energetic particle environment in the solar wind between 0.65 and 5.2~au.

On 2024 May 13, a large SEP event took place in association with an eruption close to the western limb of the Sun as seen from Earth. Providentially, at that time JUICE was close to radial alignment with the Solar Terrestrial Relations Observatory Ahead \citep[STEREO-A;][]{Kaiser2008} spacecraft, the heliocentric distance separating the two probes being only ${\sim}0.13$~au. Hence, this event is optimal not only to analyse a substantial particle event detected by multiple spacecraft in the inner heliosphere, but also to take advantage of SEP measurements from nearby locations for characterisation and cross-calibration purposes \citep[e.g.][]{Khoo2024}. In this study, we present observations and analysis of the 2024 May 13 SEP event with a particular focus on measurements from JUICE, STEREO-A, and near-Earth assets, such as the Solar and Heliospheric Observatory \citep[SOHO;][]{Domingo1995} and the Wind \citep{Ogilvie1997} spacecraft. Our main aims are to characterise JUICE observations within the interplanetary (IP) context through which SEPs propagated and to perform a cross-calibration of the energetic particle instruments on board the JUICE and STEREO-A spacecraft.

In Sect.~\ref{sec:Instrumentation} we present the spacecraft positions in the heliosphere at the time of the particle event on 2024 May 13 and list the main instrumentation used in this study. Section~\ref{sec:Solar} presents an overview of the solar eruption related to the particle event, which is discussed in detail in Sect.~\ref{sec:Analysis}.  In Sect.~\ref{sec:Discussion} we summarise and discuss the main findings of the study and in Sect.~\ref{sec:Conclusions} we outline the main conclusions.  


\section{Spacecraft positions and instrumentation} 
\label{sec:Instrumentation}

An overview of the locations where different inner heliospheric probes were situated at the onset time of the 2024 May 13 event is provided in Fig.~\ref{fig:solar_mach}(a). STEREO-A (1, red) was located at 0.96~au from the Sun and ${\sim}$13$^{\circ}$ west of Earth (2, green). JUICE (6, pink) was close to radial alignment with STEREO-A at a distance of 1.09~au. Parker Solar Probe \citep[PSP, 4, purple;][]{Fox2016} was near its aphelion, at about 0.74~au from the Sun and ${\sim}$94$^{\circ}$ west of Earth. Solar Orbiter \citep[5, blue;][]{Muller2020} was located at 0.72~au and about ${\sim}$169$^{\circ}$ west of Earth. BepiColombo \citep[3, gold;][]{Benkhoff2021} was positioned at 0.35~au between the locations of PSP and Solar Orbiter, however none of its instruments were collecting data during the SEP event investigated here. In particular, the spatial separation between JUICE and STEREO-A---0.13~au in radial distance, $0.3^{\circ}$ in latitude, and $1.6^{\circ}$ in longitude---is appropriate for a characterisation of the particle instrument on board JUICE in comparison to STEREO-A measurements.

\begin{figure*}[th!]
\centering
\includegraphics[width=.85\linewidth]{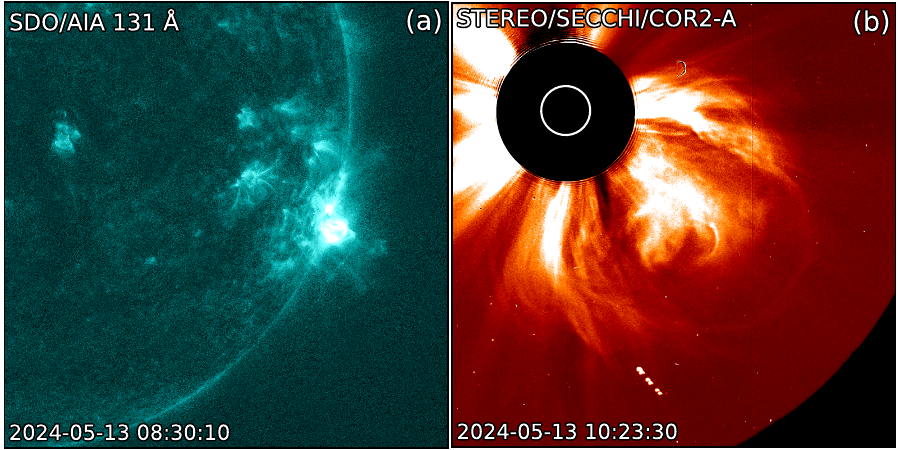}
     \caption{Overview of some of the available remote-sensing observations for the 2024 May 13 eruption. (a) SDO/AIA image in the 131~{\AA} channel showing AR~13664 and the eruption onset in the southwestern quadrant of the solar disc. (b) Coronagraph image from the COR2 telescope onboard STEREO-A displaying the CME as it propagated through the solar corona.}
     \label{fig:solar_obs}
\end{figure*}

In Fig.~\ref{fig:solar_mach}(a), each observer has been connected to the Sun via nominal Parker field lines that employ measured solar wind speeds when available. The black arrow marks the longitude of the associated flare (W81 in Stonyhurst coordinates), and the dashed black spiral depicts the nominal magnetic field line connecting to this location. For Earth, a speed of 690~km~s$^{-1}$ has been employed based on data from the Solar Wind Experiment \citep[SWE;][]{Ogilvie1995} on board the Wind spacecraft, orbiting the Sun--Earth L1 point. The field line connecting STEREO-A to the Sun assumes a solar wind speed of 700~km~s$^{-1}$ according to measurements from the Plasma and Suprathermal Ion Composition \citep[PLASTIC;][]{Galvin2008} investigation. The same value has been employed for JUICE due to its proximity to STEREO-A. For PSP, the employed solar wind speed is 530~km~s$^{-1}$ following data from the Solar Probe Cup \citep[SPC;][]{Case2020} part of the Solar Wind Electrons Alphas and Protons \citep[SWEAP;][]{Kasper2016} instrument. At Solar Orbiter, the assumed ambient wind speed is 350~km~s$^{-1}$ based on measurements from the Proton-Alpha Sensor (PAS) of the Solar Wind Analyser \citep[SWA;][]{Owen2020} suite. For BepiColombo, an average value of 400~km~s$^{-1}$ has been used due to the lack of available plasma measurements. Finally, the field line emanating from the flare location employs a wind speed of 600~km~s$^{-1}$, an approximately intermediate value between the speeds measured by STEREO-A and PSP. According to the heliospheric context depicted in Fig.~\ref{fig:solar_mach}(a), it is clear that the locations that are best-connected to the solar eruption are STEREO-A and JUICE (footpoint separation of ${\sim}30^{\circ}$), followed by Earth (${\sim}45^{\circ}$ to its east) and PSP (${\sim}50^{\circ}$ to its west), and ultimately by BepiColombo and Solar Orbiter, the latter displaying a footpoint separation of ${\sim}140^{\circ}$.

Figure~\ref{fig:solar_mach}(b) shows the corresponding ${\sim}22$~MeV proton intensities observed by different spacecraft. It employs data from JUICE/RADEM, detailed below, as well as from the Energetic and Relativistic Nuclei and Electron \citep[ERNE;][]{Torsti1995} instrument on board SOHO, orbiting the Sun--Earth L1 point, the High Energy Telescope \citep[HET;][]{vonRosenvinge2008} part of the In situ Measurements of Particles And CME Transients \citep[IMPACT;][]{Luhmann2008} suite on board STEREO-A, the Energetic Particle Instrument-High \citep[EPI-Hi;][]{Wiedenbeck2017} part of the Integrated Science Investigation of the Sun \citep[IS$\odot$IS;][]{McComas2016} on board PSP, and the High-Energy Telescope (HET) part of the Energetic Particle Detector \citep[EPD;][]{Rodriguez-Pacheco2020} on board Solar Orbiter. The plot shows how the event features, such as flux--time profiles, onset times, and peak intensities vary across the different observers. STEREO-A and JUICE observed rapidly-increasing fluxes, in agreement with their good  connectivity to the eruption. Earth and PSP observed a more gradual increase in the proton flux probably due to their larger footpoint separation to the eruption location. Solar Orbiter detected a slow and modest rise in proton fluxes, consistent with its relatively poor magnetic connection to the flare’s origin. This analysis concentrates on observations near 1 AU—specifically from JUICE, near-Earth spacecraft such as Wind and SOHO, and STEREO-A—as these three locations had better magnetic connectivity to the solar event’s source region. Hence, PSP and Solar Orbiter data are not included in the detailed analysis presented in this study. 

Additional observations of magnetic field, plasma, and particles used in this study to provide context to the aforementioned data set  are given by the Magnetic Field Investigation \citep[MFI;][]{Lepping1995}, SWE, and the Three-Dimensional Plasma and Energetic Particle Investigation \citep[3DP;][]{Lin1995} on board Wind, the Electron, Proton, and Alpha Monitor \citep[EPAM;][]{Gold1998} on board the Advanced Composition Explorer \citep[ACE;][]{Stone1998},  the Electron Proton Helium Instrument (EPHIN) part of the Comprehensive Suprathermal and Energetic Particle Analyser \citep[COSTEP;][]{Muller-Mellin1995} on board SOHO, the Magnetic Field Experiment \citep[MFE;][]{Acuna2008}, PLASTIC, the Solar Wind Electron Analyzer \citep[SWEA;][]{Sauvaud2008}, and the Low Energy Telescope \citep[LET;][]{Mewaldt2008} on board STEREO-A. The Particle Environment Package (PEP) and magnetometer (MAG) onboard JUICE remain inactive during the cruise phase—except for scheduled check-out windows and Earth gravity assist manoeuvres—until six months prior to Jupiter orbit insertion. 

\subsection{The RADEM instrument on board JUICE}
\label{sec:RADEM_instrument}

The RADEM instrument on board JUICE has been measuring high-energy electrons and protons since September 2023.  It is composed of four detectors heads, the Proton Detector Head (PDH), the Electron Detector Head (EDH), the Heavy Ion Detector Head (HIDH), and the Directional Detector Head (DDH) \citep{PintoThesis2019}. Due to the configuration of its front-end electronics, at the time of the SEP event on 2024 May 13, the EDH and DDH measured both electrons and protons, while the HIDH measured both protons and other (heavy) ions. The PDH was able to measure protons with low contamination from electrons. Therefore, in this analysis, we focus only on PDH measurements. We note that all channel configurations were changed in July 2024 to improve particle and energy discrimination. Therefore, a similar analysis approach of other SEP events would only be valid before that date and with the conditions explained below.

The PDH instrument is an eight-sensor silicon stack detector pointing anti-sunward, away from the JUICE--Sun line of sight, with a $20^{\circ}$ field of view. We note that the RADEM instrument is oriented away from the Sun, due to the spacecraft thermal constraints during the cruise phase and its location in the +X panel of the spacecraft.
At the time of the SEP event onset, the PDH was working in single coincidence mode, meaning that each detector worked independently. This means that all eight sensors (channels) measured protons with energies above a threshold energy, as described in \citep{PintoThesis2019}. Moreover, protons above 70~MeV are capable of penetrating the walls of the PDH collimator and reach the sensors. Therefore, each PDH channel alone was not capable of discriminating proton energies, working essentially as an integral energy channel. However, by linearly combining channels, as summarized in Col.~1 of Table~\ref{table:RADEMchannels}, and applying a bow-tie method \citep[e.g.][]{Raukunen2020}, summarised below, we were able to generate four differential energy proton channels. These channels were chosen based on their respective signal-to-background ratio. Since the second proton detector threshold was set very low, it had a large background, most likely due to electronic noise. To decrease the effect of background counts, we subtracted the average count rate of each channel from a quiet period, namely the day of 2024 May 6.

Since the response to protons above 70 MeV is different for each proton bin, we applied the bow-tie method using the channel response functions up to this energy only (70~MeV). The response functions were derived using the GEometry ANd Tracking (GEANT4) simulation toolkit \citep[][]{Geant4Allison}, with simulation parameters as detailed in \cite{PintoThesis2019}. The observed counting rate of a detector can be approximated as $R=j(\mathrm{Eff})\cdot G \cdot dE$, where Eff is the effective energy, G is the channel geometric factor and dE is the channel width. If we assume the flux spectrum to be a power law of energy, the bow-tie analysis method can find a unique solution for the Eff and $G \cdot dE$, independent of the spectral index of the SEP event. The spectral indices [$-5$,$-2$] encompass the majority of SEP events. As described by \cite{Raukunen2020}, we adjusted the power-law index in 0.1 steps within the specified range and derived a set of effective-energy-dependent $G \cdot dE$ curves that converged centrally, creating a characteristic “bow-tie” pattern \citep{1974VanAllen}. The point of convergence represents the optimal values for both the effective energy (Eff) and $G \cdot dE$ within the range of power-law spectra considered. This optimal point is identified by minimizing the spread between the 95th percentile ($\delta^+_G$) 
and the 5th percentile ($\delta^-_G$) of the 
$G \cdot dE$ values.

We note that this reconstruction is only valid for SEPs with negligible proton fluxes above 70 MeV, and for observations made between 2023 September and 2024 July 10, as discussed above. Additionally, while this method finds a unique value of $G \cdot dE$ to reconstruct the flux independently of the spectral index, the response of each channel still depends on it. Therefore, the quality of the reconstruction is influenced by the spectral shape at each moment in time. Table~\ref{table:RADEMchannels} shows the results using the bow-tie method, showing the four resulting channels (Col.~1), the energy ranges (Col.~2), the effective energies (Eff, Col.~3), the mean geometric factor ($G \cdot dE$, Col.~4), as well as 5th ($\delta^-_G$, Col.~5), and 95th ($\delta^+_G$, Col.~6) percentiles of the geometric factor.
The quality of the data reconstruction is better for the lowest (6.9~MeV) and highest (31.2~MeV) energy channels, which have the lowest geometric factor dispersion in relation to the spectral index, as can be seen in Cols.~5 and 6 of Table~\ref{table:RADEMchannels}.


\section{Overview of the solar eruption} \label{sec:Solar}

As described in detail by, for example, \citet{Liu2024} and \citet{Weiler2025}, the Sun exhibited considerable activity in 2024 May. A large and complex National Oceanic and Atmospheric Administration (NOAA) active region (AR) 13664 appeared from the eastern limb as seen from Earth on April 30 and disappeared behind the western limb on May 13. As it rotated with the Sun, AR 13364 produced a series of M- and X-class solar flares and CMEs. The SEP event under analysis is related to an M6.6 flare and associated CME erupting on the western limb at the beginning of day May 13. We summarise here the relevant information concerning this eruption, and an overview of some available remote-sensing observations is provided in Fig.~\ref{fig:solar_obs}. 

\subsection{The flare } \label{subsec:flare}

The source region---AR~13664, located at S17W81 in Stonyhurst coordinates---and the onset of the 2024 May 13 eruption as observed from Earth orbit by the Atmospheric Imaging Assembly \citep[AIA;][]{Lemen2012} on board the Solar Dynamics Observatory \citep[SDO;][]{Pesnell2012} in the 131~{\AA} channel are shown in Fig.~\ref{fig:solar_obs}(a). The associated M6.6-class flare was of long duration, with start time at 08:48~UT, peak time at 09:44~UT, and end time at 10:57~UT.
The GOES X-Ray flux in the top panel of Fig.~1 in \citet{Kruparova2024} shows the flux increase related to the flare under study (approximately mid day of May 13). The second and third top panels of Fig.~1 in \citet{Kruparova2024} show the Type~III radio bursts as observed by Wind/WAVES and STEREO-A/WAVES, indicating that electrons escaped from the flare eruption outwards through IP space.

\begin{figure*}[th!]
\centering
\includegraphics[width=.49\linewidth]{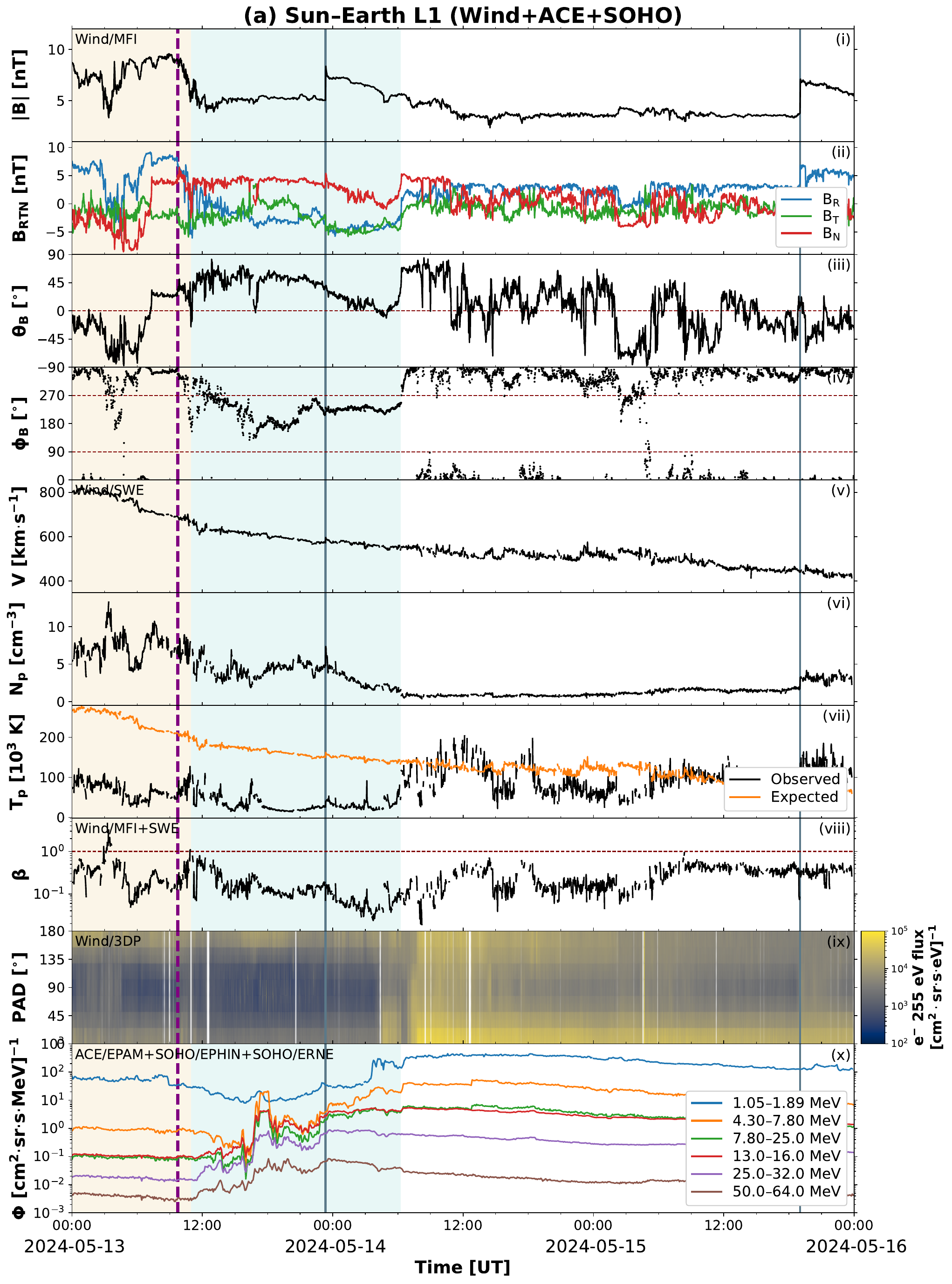}
\includegraphics[width=.49\linewidth]{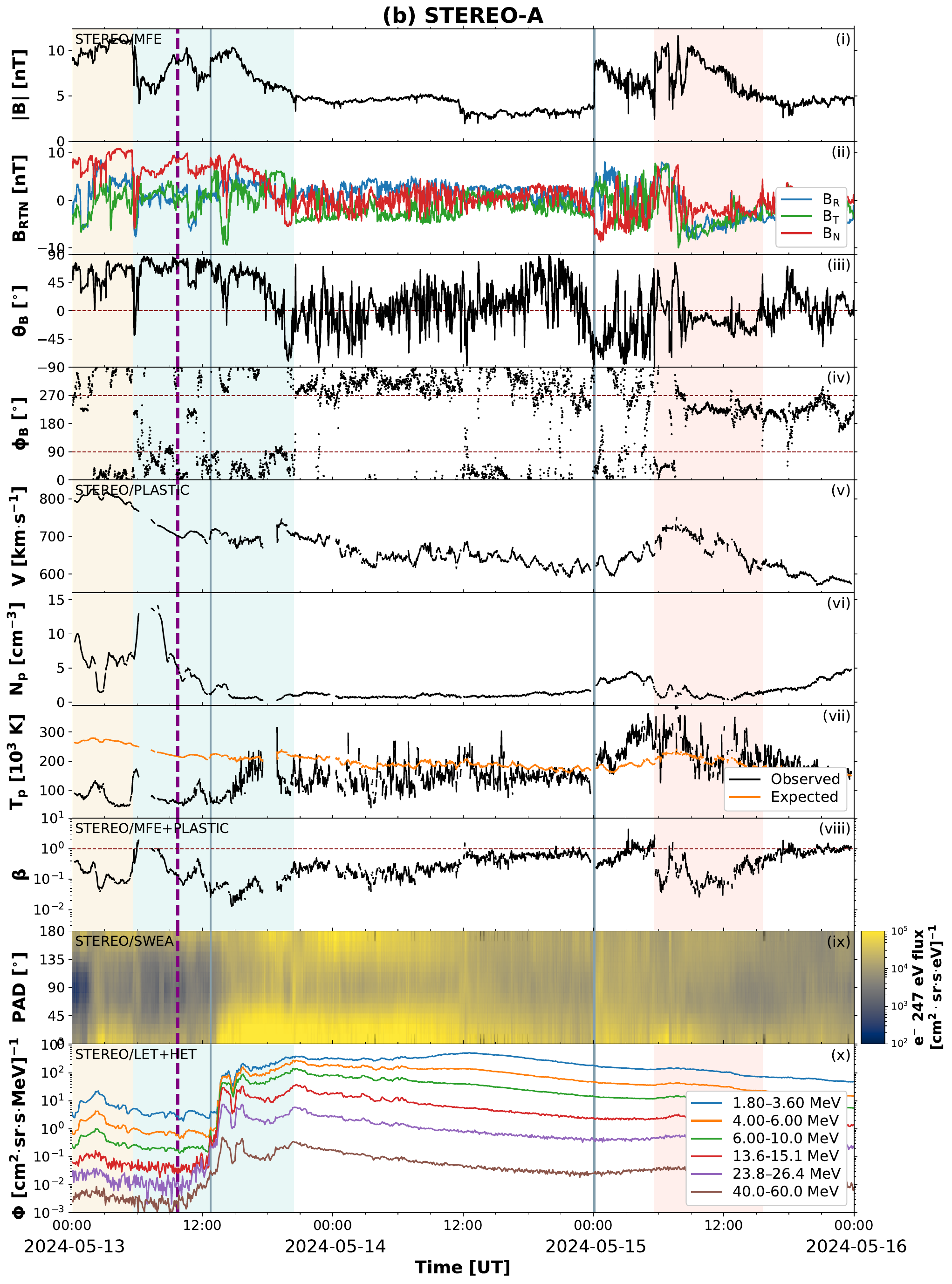}\\
\vspace*{3pt}
\includegraphics[width=.49\linewidth]{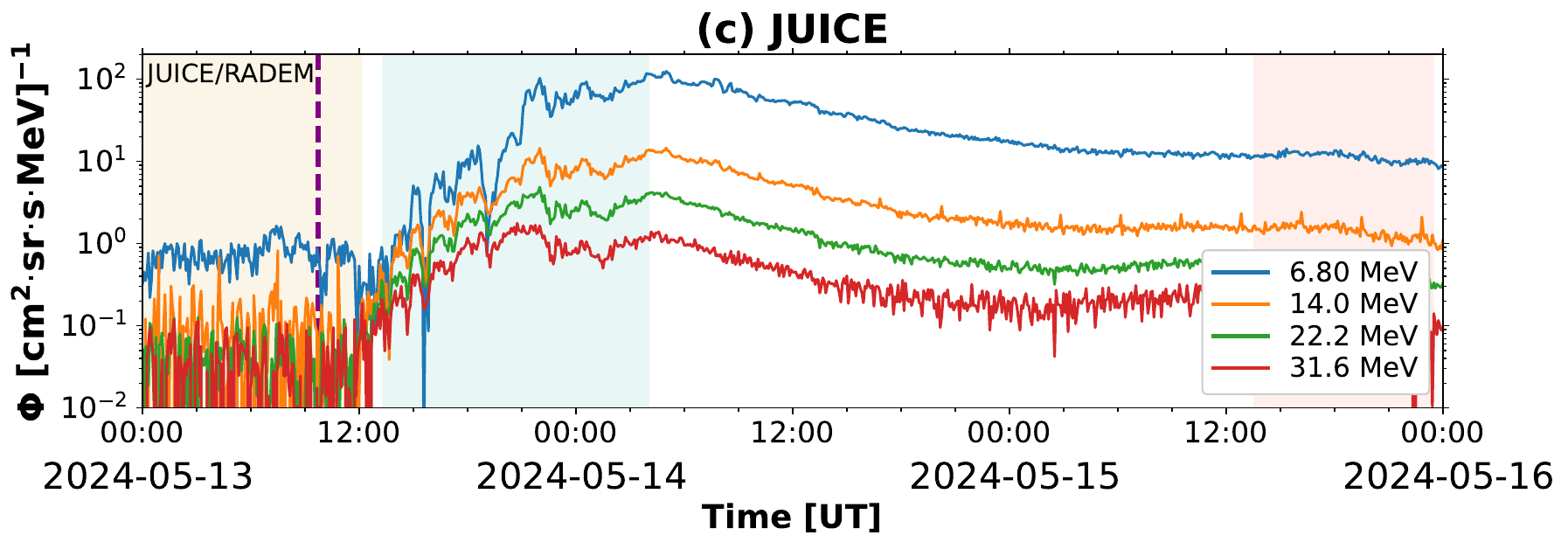}
     \caption{In-situ magnetic field and plasma observations as well as SEP time profiles by (a) near-Earth spacecraft and (b) STEREO-A, together with (c) proton observations by JUICE. The panels in (a) and (b) show, from top to bottom, the (i) magnetic field magnitude, (ii) magnetic field components \cite[where RTN stands for radial--tangential--normal coordinates; e.g.][]{Hapgood1992}, magnetic field (iii) latitudinal and (iv) azimuthal angles, $\theta$\textsubscript{B-RTN} and $\phi$\textsubscript{B-RTN}, (v) solar wind speed, (vi) proton density, (vii) proton temperature, (viii) plasma $\beta$, (ix) solar wind suprathermal electron PADs, and (x) energetic proton temporal profiles. The purple dashed line indicates the flare peak time (2024 May 13 at 09:44~UT). Solid gray vertical lines indicate the passage of IP shocks or shock-like structures, whilst shaded areas in alternating colours indicate magnetic ejecta. The panel in (c) shows the energetic proton temporal profiles (i) for JUICE, with the three marked ejecta regions being propagated from their corresponding passage times at STEREO-A. Further details are given in the main text. }
     \label{fig:ip_context}
\end{figure*}

\subsection{The coronal mass ejection} \label{subsec:CME}

Figure~\ref{fig:solar_obs}(b) shows an image from the COR2  coronagraph part of the Sun Earth Connection Coronal and Heliospheric Investigation \citep[SECCHI;][]{Howard2008} suite on board STEREO-A. It displays the CME associated to the flare eruption as it propagated through the solar corona. The CME parameters were estimated by \citet{Liu2024} using the Graduated Cylindrical Shell \citep[GCS;][]{Thernisien2006, Thernisien2009} technique, which assumes a croissant-like morphology for CMEs with two ends anchored at the Sun. The GCS model utilizes observations from multiple vantage points—specifically from STEREO-A and SOHO in this case—to reduce projection-related distortions when characterizing the CME, particularly with respect to its de-projected speed, angular width, and position in the corona.

The 3D reconstruction indicates that the CME propagates along a radial trajectory, with a Stonyhurst latitude of $-36^{\circ}$ and longitude of 85$^{\circ}$. The tilt angle ($\gamma$), which describes the orientation of the CME’s central axis relative to the solar equatorial plane, is $90^\circ$, signifying a north–south axis. The CME speed is derived from a linear fit of the leading-front distances in GCS reconstruction, giving a value of 1700~km~s$^{-1}$. The aspect ratio ($\kappa$) is 0.70 and the half angle is $25^\circ$. 
Following the approach outlined by \citet{Dumbovic2019}, the semi-angular width of the CME in the equatorial plane is determined using the formula:
${R\textsubscript{maj}-{(R\textsubscript{maj}-R\textsubscript{min})} \times |\gamma|/90}$
,
where 
$R\textsubscript{maj}$ represents the face-on half-width of the CME and is obtained by adding the half-angle to 
$R\textsubscript{min}$, the edge-on half-width. The value of 
$R\textsubscript{min}$ itself is derived using 
$\arcsin(\kappa)$. This results in a width or total angular extent of the CME of 89$^{\circ}$.  Thus, the wide CME ($\sim$89$^{\circ}$) is propagating in IP space in the direction S36W85 with a high speed ($\sim$1700~km~s\textsuperscript{-1}).


\section{Analysis of the SEP event on 2024 May 13}
\label{sec:Analysis}

In this section we first present the SEP event observed by near-Earth spacecraft (Wind, ACE, SOHO), STEREO-A, and JUICE on 2024 May 13, together with the IP context through which particles were accelerated and spread. We then proceed with in-depth analyses of the SEP pitch-angle distributions (PADs) and of the proton energy spectra from the three locations. Finally, we present a detailed comparison of JUICE, SOHO, and STEREO-A observations, including proton spectra analyses. 


\subsection{Solar energetic particle measurements and IP context}
\label{subsec:SEP_measurements}

As mentioned in Sect.~\ref{sec:Solar}, the period of 2024 May was characterised by high levels of solar activity. In particular, a sequence of CMEs launched in close succession from AR~13664 was responsible for the largest geomagnetic storm in two decades, which took place during May 10--12 \citep[e.g.,][]{Hajra2024, Liu2024, Hayakawa2025}. The link to a simulation of the heliosphere's state during the aforementioned period, using the WSA--ENLIL+Cone model \citep[][]{Odstrcil2004}, is included in Appendix \ref{app_ENLIL}, along with a description of the model and the model's input parameters.

To evaluate the IP context through which SEPs accelerated by the May 13 eruption were spread near the locations of the Earth, JUICE, and STEREO-A, we examined, aside from particle data at the three locations of interest, also magnetic field and plasma measurements near Earth and at STEREO-A. We note that evaluating the IP status is crucial for understanding how particles reach the different spacecraft we aim to compare, as IP structures can influence their fluxes and anisotropies \citep{RichardsonCane1996, Rodriguez-Garcia2025}.   

These combined observations of magnetic field, plasma, and particles are all displayed in Fig.~\ref{fig:ip_context}. To identify the different structures impacting the two locations for which magnetic field and plasma data are available (i.e., Earth and STEREO-A), we searched for signatures indicating the passage of shocks as well as IP CMEs \citep[hereafter ICMEs, e.g.][]{Zurbuchen2006}.The so-called magnetic clouds (displaying a clear flux rope structure) are easily identifiable via ``classic'' signatures such as (1) an increase in the magnetic field strength, (2) a monotonic magnetic field rotation, (3) low proton temperature, and (4) plasma $\beta$ below 1 \citep{Burlaga1981}. The event under study features multiple instances of CME--CME interaction \citep[e.g.][]{Lugaz2017}, hence it is not straightforward to isolate individual eruptions in the in-situ time series. Nevertheless, we attempt to separate the arrivals of distinct magnetic field and plasma environments, noting that at least a portion of them may have undergone interaction and merging before reaching 1~au. We identified a (complex) ejecta when at least the following conditions were met: plasma $\beta$ below 1, a lower-than-expected temperature, rotation in the magnetic field components, and lower fluctuations in the magnetic field in comparison to the ambient solar wind. In the following, we describe in deeper detail the IP context at the three locations emerging from our analysis.

\subsubsection{Solar energetic particle observations and IP context: Earth}
\label{subsec:SEPS_IP_context_Earth}

Figure~\ref{fig:ip_context}(a) shows the magnetic field, plasma, and particle observations by near-Earth spacecraft from 2024 May 13 to May 16. The peak of the solar flare associated with the SEP event on May 13 is indicated with the vertical dashed purple line. At the time of the SEP onset, near-Earth spacecraft were embedded in a series of interacting ejecta indicated with the golden and aqua shadings from 07:20~UT on May 13 to 06:16~UT on May 14. We observed rotations in the magnetic field components (iii, iv), a low plasma beta (viii), evidence of speed expansion (v), a lower-than-expected temperature (vii), and bi-directional suprathermal electron PADs.  The bottom panel (x) shows the proton intensity profile as measured by ACE/EPAM, SOHO/EPHIN, and SOHO/ERNE, presenting a gradual rise of energetic protons above 13~MeV up to at least 50~MeV. We observe a flux peak around 18:00~UT on May 13, likely related to a sudden change in the magnetic field orientation (panel (iii)). The lower ion energies only increase in the rear part of ejecta after the passage of a shock-like wave impacting near-Earth space at 23:20~UT on May 13 indicated with the grey vertical line. Protons arrived to the spacecraft from the Sun at pitch-angle 180 (inwards polarity), and present a bidirectional flow of suprathermal electrons accompanied by a depletion at a pitch angle of 90 lasting ${\sim}12$ hours, consistent with the presence of a magnetic ejecta, as discussed in Sect.~\ref{subsec:SEP pitch-angle distributions}. Near-Earth spacecraft also observed a prior energetic storm particle (ESP) event associated to a shock arrival on May 10 and an SEP event on May 11 followed by several ESP events associated to IP shocks arriving at Earth (not shown). This period is the one related to the intense geomagnetic storm discussed by \citet{Hajra2024}, \citet{Liu2024}, \citet{Hayakawa2025}, and \citet{Weiler2025}, among others. 

\subsubsection{Solar energetic particle observations and IP context: STEREO-A}
\label{subsec:SEPS_IP_context_STA}


Figure~\ref{fig:ip_context}(b) shows the magnetic field, plasma, and particle observations by STEREO-A from 2024 May 13 to May 16.  At the time of the SEP onset, the STEREO-A spacecraft was embedded in an ejecta indicated with the aqua shading from 05:38~UT to 20:27~UT on May 13. This is the same complex ejecta identified at the near-Earth location, marked with the same colour.
The bottom panel (x) shows a clear proton event observed up to energies of ${\sim}40$~MeV with a fast increase in the measured fluxes coinciding with the passage of a shock-like wave marked with the grey vertical line (likely corresponding to the one identified at Earth), where clear velocity dispersion is also present. We observed a double peak in the flux of protons followed by a gradual increase until the end of the ejecta, which marks the start of the decay phase of the particle enhancement. The depletion of particles in between the two peaks might be related to a sudden change in the magnetic field orientation (panel (iii)). The first arriving protons reached the spacecraft from the Sun at pitch-angle 0 (outwards polarity), as discussed in Sect.~\ref{subsec:SEP pitch-angle distributions}. The prior ESP event that occurred on May 10 as well as the SEP event of May 11 are also measured by STEREO-A, with proton flux profiles that are qualitatively similar to the ones detected at near-Earth spacecraft (not shown).

\subsubsection{Solar energetic particle observations: JUICE}
\label{subsec:SEPS_IP_context_JUICE}

\begin{figure*}[htb]
\centering
  \resizebox{0.7\hsize}{!}{\includegraphics{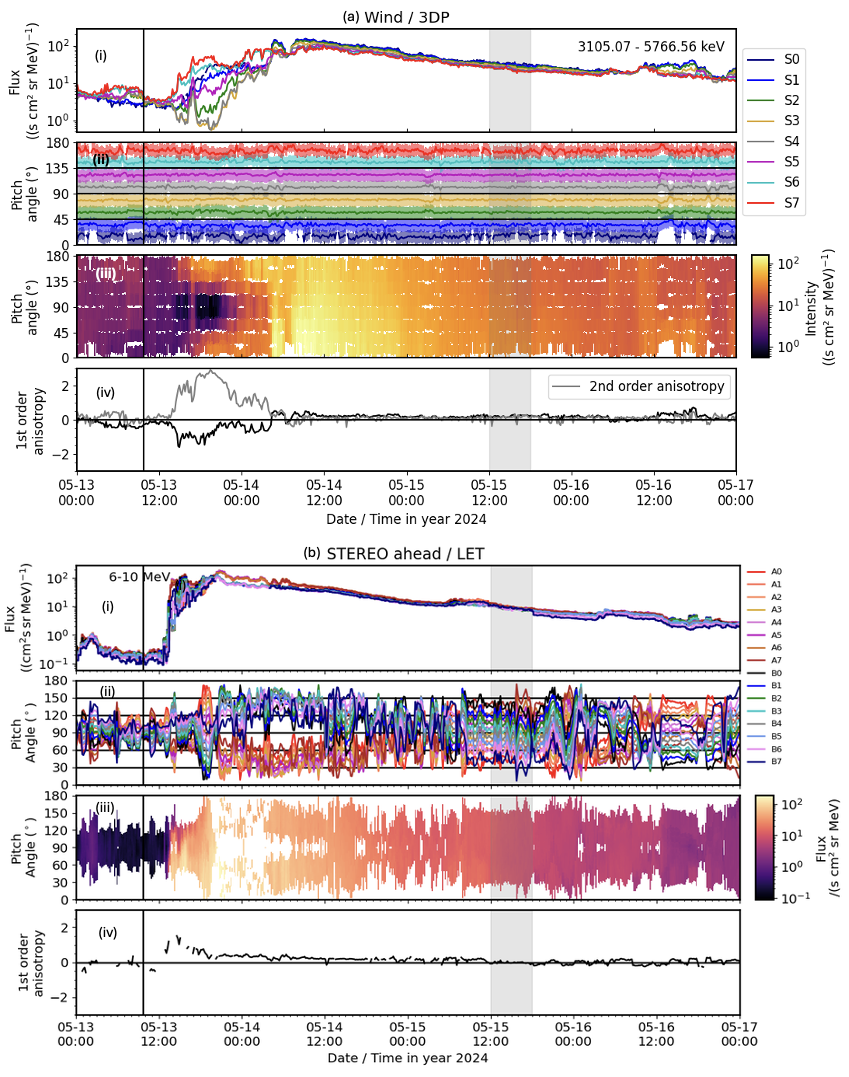}}
     \caption{Pitch-angle distributions of protons recorded by Wind/3DP at ~4 MeV (panel a) and STEREO-A/LET at ~6 MeV (panel b). The panels include: (i) the measured intensities within each instrument’s field of view; (ii) the pitch-angle coverage, displayed for the eight angular bins of Wind/3DP (a) and the central angles of 16 STEREO-A/LET sectors (b), with front-facing sectors shown in red hues and rear-facing in blue hues; (iii) a colour-coded representation of the pitch-angle intensity distribution; and (iv) first-order anisotropy values, ranging from -3 to 3, following the approach of \cite{Dresing2014}. A vertical line marks the timing of the soft X-ray flare peak (approximately 09:44 UT), which is linked to the associated SEP event. Details given in the main text.}  
     \label{fig:anistrop_protons}
\end{figure*}

\begin{figure*}[th!]
\centering
\includegraphics[width=.7\linewidth]{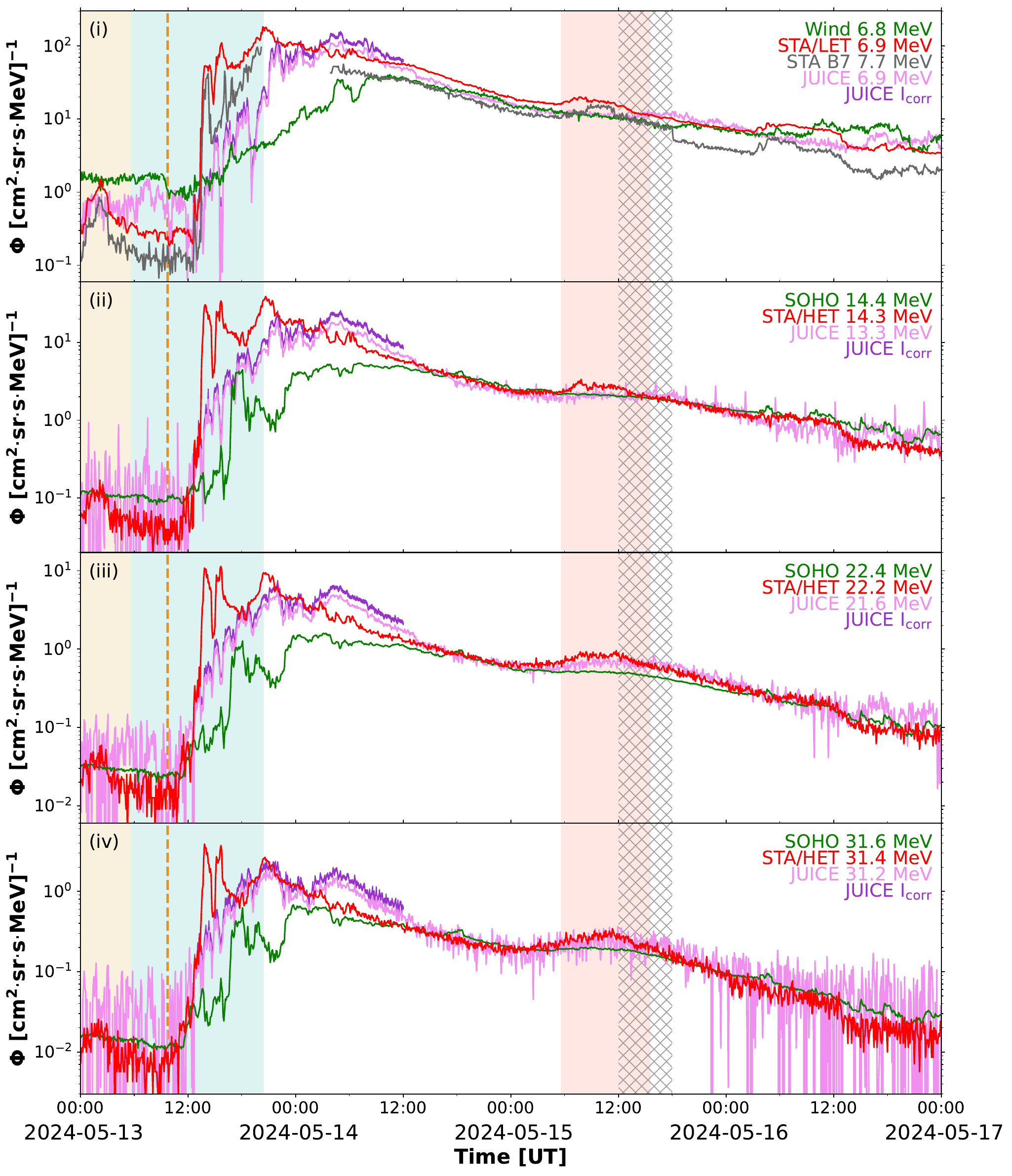}
     \caption{In-situ SEP time profiles by Wind/3DP--omni-- or SOHO/ERNE--sun-- (green), STEREO-A/LET--omni-- or HET--sun-- (red), and JUICE/RADEM--anti-sun-- (magenta) for four different proton energies: (i) ${\sim}6.9$ MeV, (ii) ${\sim}13.3$, (iii) ${\sim}21.6$ MeV and (iv) ${\sim}31.2$ MeV. JUICE $I_\mathrm{corr}$ is estimated applying a radial scaling in the fluxes to STEREO-A location. The orange dashed line in the four panels indicates the flare peak time, the various shadings mark the ejecta present at the location of STEREO-A discussed in Sect.~\ref{subsec:SEPS_IP_context_STA}, and the gray hatched area the isotropic period discussed in Sect.~\ref{subsec:SEP pitch-angle distributions}. Further details are given in the main text.  }
     \label{fig:seps_by_energy}
\end{figure*}

The bottom panel (i) of Fig.~\ref{fig:ip_context}(c) shows the SEP event on 2024 May 13 observed by JUICE/RADEM, using the four resulting channels described in Sect.~\ref{sec:RADEM_instrument}, with effective energies of 6.9~MeV (blue line), 13.3~MeV (orange line), 21.6~MeV (green line), and 31.2~MeV (red line).  The data correspond to particles coming from the anti-Sun direction, as discussed above, with a resolution of five minutes. The observed profiles show a relatively fast rise of energetic protons after the solar flare peak time that reaches energies of at least 31~MeV. The magnetometer on board JUICE was not measuring at that time, hence it is not possible to confirm the concurrent presence of an ejecta at JUICE's location.  Nevertheless, the spacecraft's proximity to STEREO-A enables us to assume that JUICE was embedded at the onset time of the May 13 SEP event in the same two interacting/merged structures indicated in Fig.~\ref{fig:ip_context}(b) by the golden and aqua shaded areas.

These structures were time-shifted to JUICE, assuming average propagation speeds of 820~km~s$^{-1}$ and 710~km~s$^{-1}$, respectively, based on PLASTIC data, which are also marked in Fig.~\ref{fig:ip_context}(c) with similar colour shading. The presence of a magnetic ejecta may affect the profile of energetic protons observed by RADEM, leading to a depletion in particle flux that coincides with the centre of the structure indicated by the aqua shading, as previously observed by STEREO-A. We have also time-shifted to JUICE the small ejecta observed at STEREO-A during May 15 (salmon-shaded region), that is, towards the decay phase of the SEP event, using an average speed of 680~km~s$^{-1}$ based again on PLASTIC data. 

RADEM also observed the previous SEP event on 2024 May 11 (not shown). However, this portion of the data cannot be reconstructed straightforwardly due to both the single coincidence mode of the PDH, which makes the sensors sensitive to penetrating particles above 70~MeV, and the large fluxes of high-energy particles related to the May 11 SEP event.  


\subsection{Solar energetic particle pitch-angle distributions}
\label{subsec:SEP pitch-angle distributions}

In this section we study the PAD of STEREO-A and Wind, both of which provide energetic particle anisotropy information. Due to the different pointing directions of STEREO-A and SOHO compared to JUICE, it is necessary to analyse the PADs to identify periods of isotropy for meaningful comparison of particle fluxes. We used the 16 viewing directions of the LET instrument on board STEREO-A. It is important to note that the pitch-angle coverage provided by STEREO-A/LET is influenced by the alignment between the magnetic field and the instrument’s viewing geometry. In contrast, the Wind spacecraft, which is spin-stabilized and equipped with a wide-field telescope, enables the 3DP instrument to sample a broader portion of the sky. This design allows for a more comprehensive reconstruction of the three-dimensional particle distribution.

\subsubsection{Solar energetic particle pitch-angle distributions: Earth}
\label{subsubsec:subsubsec:SEP pitch-angle distributions_near_Earth}

Figure~\ref{fig:anistrop_protons}(a) shows the PAD of protons measured by Wind/3DP at 4~MeV. Panel (i) shows the intensities observed by each pitch-angle bin, while panel (ii) shows the pitch-angle coverage of each of the eight publicly available pitch-angle bins of the instrument. Panel (iii) presents the PAD with colour-coded intensities and panel (iv) indicates the first-order anisotropy, in the range [$-3$, 3] \citep[e.g.][]{Dresing2014} and the second-order anisotropy. We note that periods are considered isotropic when the first and second-order anisotropy is low, $\lesssim$|1|. This panel shows that the early phase of the ${\sim}3.1$--5.7~MeV proton event is anisotropic for more than twelve hours, showing also a strong bidirectional component. From the onset of the SEP event---shortly after the soft-X ray peak of the flare indicated with the vertical line---until 06:00 UT on May 14 we observe higher fluxes in the bins measuring particles coming from the Sun (panel (i)) that corresponds to pitch angles near 180$^{\circ}$ (panel (ii)), consistent with the local negative magnetic polarity shown in Fig.~\ref{fig:ip_context}(a). Starting around midday on May 13 and lasting for approximately twelve hours, we observe a bidirectional flow along the magnetic field (pitch angles of 0$^{\circ}$ and 180$^{\circ}$) with a depletion in intensity around pitch angles of 90$^{\circ}$. This feature is characterized by the large second-order anisotropy (panel iii) and is consistent with the presence of a magnetic ejecta, as discussed in Sect. \ref{subsec:SEPS_IP_context_Earth}. After this period, the first-order anisotropy becomes positive, consistent with the sunward-looking bins observing higher fluxes of particles with pitch angles near 0$^{\circ}$. During the decay phase of the SEP event, from May 15 to May 16 the flux becomes isotropic, namely the first-order anisotropy is $\sim$0. We indicate with the grey shading in Fig.~\ref{fig:anistrop_protons}(a), the period selected for the intercalibration analysis, as discussed below.  

\subsubsection{Solar energetic particle pitch-angle distributions: STEREO-A}
\label{subsubsec:SEP pitch-angle distributions_STEREO-A}

Panel (b) of Fig.~\ref{fig:anistrop_protons} displays proton intensities in the 6–10 MeV range as measured by STEREO-A/LET across its 16 sectors, with eight forward-facing sectors represented in shades of red and eight rear-facing sectors shown in shades of blue. LET measured an eighteen-hour anisotropic period starting shortly after 12:00~UT on May 13, where most of the particles are observed in the sunward-facing sectors.  The vertical line indicates the soft-x ray flare peak time. 
The pitch-angle coverage is not good during the SEP onset period, only covering $60-120^{\circ}$ as shown in panel (ii), which presents the pitch-angles of the sector centres. During the decay phase of the SEP event, from May 15 to May 17, the flux becomes isotropic. We selected the period from May 15, 12:00 to 18:00~UT as the time with the lowest first-order anisotropy value, marked by the grey shading in the figure, for use in the intercalibration analysis, as discussed in Sect.~\ref{subsec:proton_spectra}. We note that, in selecting the grey-shaded area, we intentionally excluded the period from May 15, 18:00~UT to May 16, 02:00~UT, because, although the anisotropy remains very low, the pitch-angle coverage changes, as shown in panel (ii).

\subsection{Comparison between JUICE, STEREO-A, and near-Earth measurements}
\label{subsec:JUICE_STA_comparison}

Figure~\ref{fig:seps_by_energy} shows the time profiles for each of the four effective energies, namely 6.9, 13.3, 21.6, and 31.2~MeV measured by JUICE (magenta) and the correspondent channels at STEREO-A (red) and near-Earth (green), as shown in the legend. STEREO-A presents the most prompt increase in the signal of protons reaching the peak intensity shortly after the SEP event onset. We note that the presence of the magnetic ejecta, indicated by the aqua-shaded area marking its arrival at STEREO-A, modulated the particle profile, causing rapid decreases and increases in their fluxes, as discussed in Sect.~\ref{subsec:SEPS_IP_context_STA}. The JUICE spacecraft also measures similar peak proton fluxes but with a slower increase, probably related to the anti-Sun--spacecraft-line pointing of the instrument. Wind and SOHO show a slower increase and lower peak intensities that might be related to their worse magnetic connectivity as discussed in Sect.~\ref{sec:Instrumentation}. 

The main difference in the flux of particles between STEREO-A/LET and HET, and JUICE/RADEM, as measured by the four proton channels of Fig.~\ref{fig:seps_by_energy}, occurs during the SEP onset, namely between the flare peak time indicated with the dash vertical line and middle of day 2024 May 14. This disparity might be partly related to the difference in the field of view, as the period mentioned above is anisotropic as discussed in Sect.~\ref{subsec:SEP pitch-angle distributions}.  We note again that JUICE/RADEM is looking away from the Sun in the radial direction while STEREO-A/LET fluxes are summed along the 16 sectors and STEREO-A/HET HET is viewing along the nominal Parker spiral.

We also show in the first panel of Fig.~\ref{fig:seps_by_energy} the fluxes measured by STEREO-A/LET (gray line) only in the sector B7, namely looking against the direction to the Sun with a field of view of $25^{\circ}$, similar to the JUICE/RADEM field of view. We note that this STEREO-A/LET sectored data set is only available for the ${\sim}$7.7~MeV proton energy channel. In Appendix~\ref{app:data_gap} we explain the reason behind the data gap shown in the B7 sectored data. We observed a dissimilarity in the fluxes measured by JUICE (magenta line) in comparison to STEREO-A/LET/B7 (grey line). Although it could be related to not being exactly observing the same field of view, the fact that the internal structures of the ejecta pass by STEREO-A and JUICE at different times, potentially modulating the particle profile differently, could be an important factor. 

With a lower impact, the difference in radial distance between STEREO-A and JUICE (0.13~au) might be also affecting the measured fluxes of particles. As demonstrated by, for example, \citet{Lario2006} and \citet{Rodriguez-Garcia2023a}, SEP intensities tend to depend on heliocentric distance. In order to evaluate this effect on the difference in the flux measured by JUICE,  we applied a radial scaling in the fluxes. Figure~\ref{fig:seps_by_energy} shows the JUICE intensity values, corrected ($I_\mathrm{corr}$) for radial distance, as indicated by the purple-shaded curve. We scaled it to the location of STEREO-A, with corrected peak intensities calculated as a radial dependence of  $\sim R^\alpha$ as detailed by  \citet{Lario2006}, namely $\alpha$=log($I_\mathrm{observed}$/$I_\mathrm{corrected}$)/log($R_\mathrm{observed}$/$R_\mathrm{corrected}$),  where $\alpha = { a \pm b}$, and $a=2.14$, $b=0.26$ for 4--13~MeV protons, and  $a=1.97$, $b=0.27$ for 27--37~MeV protons. As illustrated in the figure, the the radial correction is small compared to the differences between the STEREO-A and JUICE intensities. 
  
We indicate in Fig.~\ref{fig:seps_by_energy} with a grey hatched area the isotropic period found in Sect.~\ref{subsec:SEP pitch-angle distributions}. This period partially coincides with an ejecta structure arriving at STEREO-A, indicated with the salmon shading, which apparently did not influence the isotropisation of the proton fluxes. The three locations---near-Earth assets, STEREO-A, and JUICE---exhibit very similar proton temporal profiles, as the particles may be uniformly distributed in longitude and radial distance within the heliosphere due to the reservoir effect, with comparable intensities observed between the distant spacecraft \citep{1972McKibben, Roelof1992, 2010Lario}. 

\begin{figure}[th!]
\centering
\includegraphics[width=0.99\linewidth]{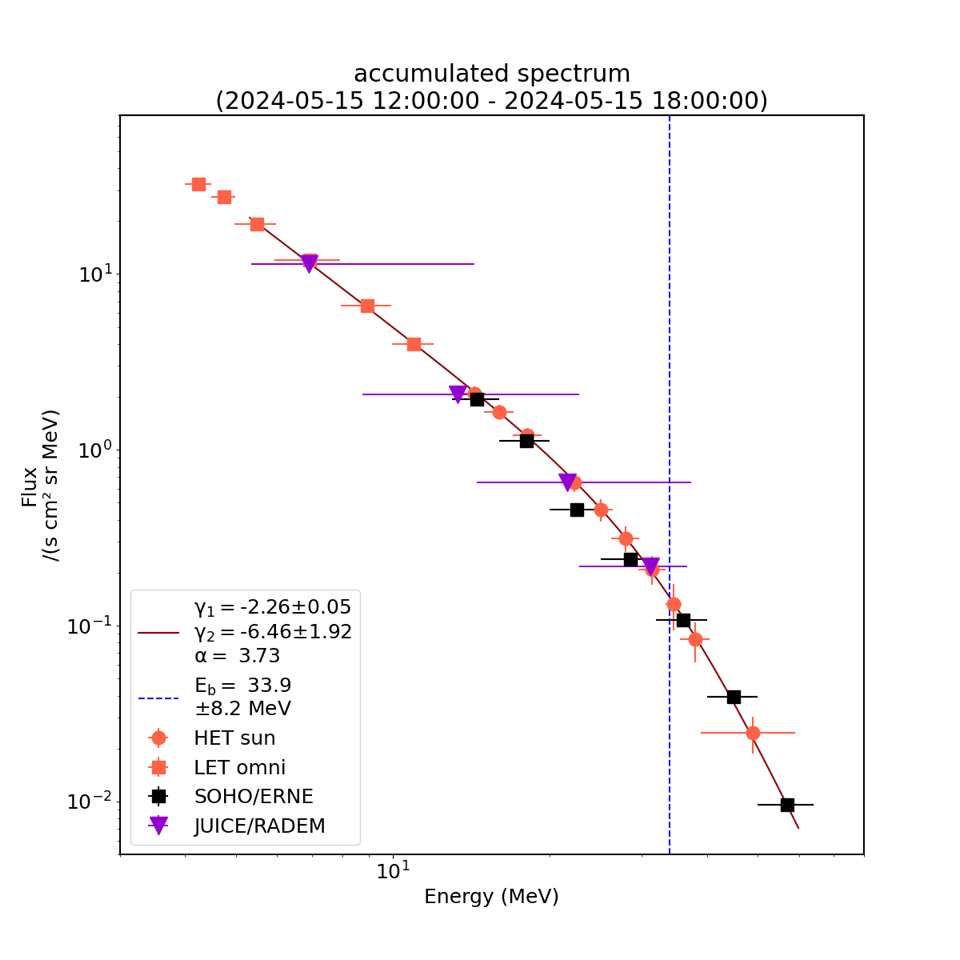}
     \caption{Proton accumulated spectra measured by STEREO-A (red), SOHO/ERNE (black), and JUICE/RADEM (magenta). The legend shows the fit values using STEREO-A data: the spectral index ($\delta$\textsubscript{1}, $\delta$\textsubscript{2}) observed in between the spectral transition: E\textsubscript{b}; and $\alpha$, which determines the sharpness of the break \citep{Strauss2020}. Details given in the main text.}
     \label{fig:proton_spectra}
\end{figure}

\subsection{Energy proton spectra and intercalibration analysis}
\label{subsec:proton_spectra}

 To perform a cross-calibration of the energetic particle instruments on board the JUICE and STEREO-A spacecraft, we determined the proton spectra, as observed by STEREO-A during the isotropic period. This interval spans from May 15, 2024, 12:00 to 18:00~UT, as indicated by the grey shading in Figs.~\ref{fig:anistrop_protons} and \ref{fig:seps_by_energy} (hatched). We used the omnidirectional data from STEREO-A/LET and HET (Sun-directed) and following the method described by \citet{Dresing2020} and \citet{Strauss2020} we fit the spectrum using Orthogonal Distance Regression \citep[ODR; ][]{Boggs1990} provided by the SciPy Python package \citep{Virtanen2020}. We note that, different to the approach by \citet{Dresing2020} and \citet{Strauss2020}, we used here the errors of the fit parameters as returned by ODR, which represent the standard deviations of the estimated parameters. We found a broken power law to best describe the data represented by
\begin{equation}
    I(E) = I_0 \left( \frac{E}{E_0}\right)^{\delta_1} \left( \frac{E^\alpha+E^\alpha_b}{E^\alpha_0+E^\alpha_b}\right)^\frac{\delta_2-\delta_1}{\alpha}.
\end{equation} 
\noindent
This model yields a spectral transition at the energy $E_b$, where $\delta_1$ and $\delta_2$ are the spectral indices at energies below and above $E_b$. The parameter $\alpha$ describes the sharpness of the spectral transition and $E_0$ is a reference energy at 0.1 MeV.
The results are shown in Fig.~\ref{fig:proton_spectra}, where STEREO-A/LET and HET are shown with the reddish points. The spectral index below (above) the spectral transition is $\delta_1=-2.26\pm0.05$ ($\delta_2=-6.46\pm1.92$), where the spectral break/transition energy is $E_b=33.9 \pm 8.2$~MeV. We note the high uncertainty in the second spectral index. Modifying the time period chosen for the accumulated spectrum (not shown) alters the spectral transition energy, potentially influencing the intercalibration factor for the fourth JUICE energy channel. 

We also show in Fig.~\ref{fig:proton_spectra} the accumulated spectrum as measured by SOHO/ERNE (black points). We note that five out of seven points almost perfectly agree with the fit based on STEREO-A data. For the determination of an intercalibration factor for JUICE/RADEM, we used the spectral fit from STEREO-A as the reference for the particle environment measured by JUICE and estimated the ratios in the proton flux measured by JUICE (magenta points) compared to the fit. We did not apply any radial/longitudinal scaling for the spacecraft measurements based on the reservoir effect discussed in Sect.~\ref{subsec:JUICE_STA_comparison}. Table~\ref{table:intercalibration} shows the derived intercalibration factors of each of the effective energy channels of JUICE/RADEM. The proton fluxes measured at the effective energies of 6.9, 13.3, 21.6, and 31.2~MeV by the radiation monitor onboard JUICE agree with the STEREO-A measurements, with a deviation of less than 25\%. We note that the smallest intercalibration factors, specifically 1.02 and 0.95, are obtained for the first (6.9~MeV) and last (31.2~MeV) channel, respectively, for which the quality of the data reconstruction, as described in Sect.~\ref{sec:RADEM_instrument}, is better. 

\begin{table}
\centering
\caption{Intercalibration factors between STEREO-A and JUICE.}
\begin{tabular}{cc}
\hline
JUICE Eff (range, MeV) &
Intensity ratio (STA/JUICE) \\ \hline
 6.9 (5.35-14.35) & 1.02                     \\
 13.3 (8.75-22.75) & 1.23                   \\
 21.6 (14.50-37.37) & 1.12                  \\
 31.2 (22.80-36.60)   & 0.95  \\\hline

\label{table:intercalibration}
\end{tabular}
\end{table}
\section{Summary and discussion}
\label{sec:Discussion}

JUICE was launched in April 2023, and it is now in its cruise phase to Jupiter, where it is scheduled to arrive in July 2031. JUICE carries RADEM, a radiation monitor that operates continuously to measure protons, electrons, and ions, pointing mainly in the anti-Sun--spacecraft direction. One of its instruments, the PDH is an eight-sensor silicon stack detector with a field of view of $20^{\circ}$, which was configured from September 2023 to July 2024 in a single-coincidence mode, meaning that each sensor worked independently. As a result of linearly combining bins and applying a bow-tie method described in Sect.~\ref{sec:RADEM_instrument}, we derived four differential channels with effective energies of 6.9, 13.3, 21.6, and 31.2~MeV, as summarised in Table~\ref{table:RADEMchannels}.

During 2024 May, AR~13664 produced a series of CMEs and associated SEP events. This period is the one related to the intense geomagnetic storm discussed by \citet{Hajra2024}, \citet{Liu2024}, \citet{Hayakawa2025}, and \citet{Weiler2025}, among others. On 2024 May 13, a large SEP event took place in association with an eruption close to the western limb of the Sun as seen from Earth (Fig.~\ref{fig:solar_obs}). Several spacecraft in the heliosphere observed the SEP event, including JUICE, STEREO-A, near-Earth spacecraft (Wind, ACE, SOHO), PSP, and Solar Orbiter, as shown in the proton flux profiles in Fig.~\ref{fig:solar_mach}b. Providentially, at that time JUICE was located very close to STEREO-A, with a difference of 0.13~au in radial distance, $0.3^{\circ}$ in latitude, and $1.6^{\circ}$ in longitude, as shown in Fig.~\ref{fig:solar_mach}a.  Therefore, in this study we aimed to characterise JUICE observations and perform a cross-calibration of the energetic particle instruments aboard the JUICE and STEREO-A spacecraft.

For this purpose, we focused our analysis on spacecraft located near 1~au, namely JUICE, STEREO-A, and near-Earth assets, all of which were well connected to the parent solar source, as shown in Fig.~\ref{fig:solar_mach}a,  using the nominal Parker spirals. To evaluate the IP context through which SEPs accelerated by the May 13 eruption were spread, we examined, in addition to particle data from the three locations of interest, magnetic field and plasma measurements near Earth and at STEREO-A, as shown in Fig.~\ref{fig:ip_context}. At the time of the SEP event, STEREO-A was embedded in a magnetic ejecta, which was likely also present at JUICE's location, since the magnetometer on JUICE was not operating to confirm this. 

We studied the proton anisotropies measured by Wind and STEREO-A. We found anisotropic periods lasting a few hours during the SEP onset, which evolved into an isotropic period during the decay phase of the event, as shown in Fig.~\ref{fig:anistrop_protons}. We selected the period from May 15, 12:00 to 18:00 hours as the interval with the lowest first-order anisotropy at the location of STEREO-A, indicated by the gray shading in the Fig. ~\ref{fig:anistrop_protons}.
In Fig.~\ref{fig:seps_by_energy}, we compared the proton flux observations of the three selected spacecraft for similar energies to the four effective channels of JUICE. We observed dissimilarities at the onset of the SEP event due to several factors: the different fields of view of the spacecraft, the local magnetic ejecta, and, to a lesser extent, the varying radial distances. However, during the period indicated by the grey hatched area in Fig.~\ref{fig:seps_by_energy}, corresponding to the selected isotropic period, the three spacecraft exhibit similar proton fluxes. 

We therefore considered the isotropic period during the decay phase of the SEP event to compute the accumulated proton spectrum for the three spacecraft. We note that the selection of this isotropic period for comparison with JUICE is particularly important, as RADEM primarily observes particles coming from the anti-Sun direction. We fitted the STEREO-A spectra with a double power law, as shown in Fig.~\ref{fig:proton_spectra}. The results of the fit showed no significant variation within the sub-periods of the selected period (not shown), except for the second spectral index, which exhibited larger uncertainty. The different channels of the near-Earth proton spectra agreed closely with the STEREO-A fit. 

We considered STEREO-A spectral fit as the particle environment measured by JUICE and estimated the differences in the proton flux measured by JUICE in comparison with the fit. Based on the results shown in Fig.~\ref{fig:proton_spectra} and summarised in Table~\ref{table:intercalibration}, the proton fluxes measured at the effective energies of 6.9, 13.3, 21.6, and 31.2~MeV by the radiation monitor onboard JUICE, agree  with a deviation less than 25\% with respect to STEREO-A measurements. 

We note that the results obtained in this study are valid only for RADEM data collected before 2024 July 10, when the instrument’s configuration was changed. We also note that the method used here for the intensity comparison carries some uncertainties. While this method finds a unique value of $G\cdot dE$ to reconstruct the RADEM flux independently of the spectral index, the response of each channel still depends on it. Therefore, the quality of the RADEM intensity reconstruction is influenced by the spectral shape at each moment in time.  
Future work will examine other SEP events with different particle spectral shapes to compare the intercalibration results. 


\section{Conclusions}
\label{sec:Conclusions}
This work illustrates that The RADEM instrument aboard JUICE is a valuable tool for measuring SEP events in the heliosphere, providing an excellent opportunity to study and characterise the energetic particle environment in the solar wind between 0.65 and 5.2~au. The proton fluxes measured at the effective energies of 6.9, 13.3, 21.6, and 31.2~MeV by RADEM, agree,  with a deviation less than 25\%, with STEREO-A measurements, with intercalibration factors between STEREO-A and JUICE of 1.02, 1.23, 1.12, and 0.95, respectively. This result is valid only for RADEM data collected before 2024 July 10.


\begin{acknowledgements}

L.R.-G.\ acknowledges support through the European Space Agency (ESA) research fellowship programme.
L.R.-G.\ and E.P.\ acknowledge support from ESA through the Science Faculty -- Funding reference ESA-SCI-E-LE-050.
E.P.\ also acknowledges support from NASA's HGI-O (no.\ 80NSSC23K0447), LWS (no.\ 80NSSC19K0067), and LWS-SC (no.\ 80NSSC22K0893) programmes.
N.D.\ and J.G.\ acknowledge funding from the European Union's Horizon Europe research and innovation programme under grant agreement No.\ 101134999 (SOLER). The paper reflects only the authors' view and the European Commission is not responsible for any use that may be made of the information it contains.
Work in the University of Turku was performed under the umbrella of Finnish Centre of Excellence in Research of Sustainable Space (FORESAIL) funded by the Research Council of Finland (grant No.\ 352847). N.D.\ is grateful for support by the Research Council of Finland (SHOCKSEE, grant No.\ 346902). We also thank the members of the Data Analysis Working Group at the Space Research Laboratory of the University of Turku, Finland for useful discussions.

The UAH team acknowledges the finnancial support of MICIU/AEI/10.13039/501100011033 and FEDER, UE through project PID2023-150952OB-I00.

C.M.S.C.\ acknowledges additional partial funding from NASA grants 80NSSC22K0893, 80NSSC21K1327, 80NSSC20K1815, and 80NSSC19K0067.

The authors thank NASA's Community Coordinated Modeling Center (CCMC; \href{https://ccmc.gsfc.nasa.gov}{https://ccmc.gsfc.nasa.gov}) for supporting the WSA--ENLIL+Cone simulation efforts presented in this work. The WSA model was developed by C.~N.~Arge (currently at NASA Goddard Space Flight Center) and the ENLIL model was developed by D.~Odstrcil (currently at George Mason University).
The WSA--ENLIL+Cone simulation run employed in this work can be accessed online at \href{https://ccmc.gsfc.nasa.gov/ungrouped/SH/Helio_main.php}{https://ccmc.gsfc.nasa.gov/ungrouped/SH/Helio\_main.php} (run id: \textsl{Laura\_Rodriguez-Garcia\_092424\_SH\_1}).

\end{acknowledgements}

\begin{flushleft}

\textbf{ORCID iDs} 
\vspace{2mm}

Laura Rodríguez-García \orcid{https://orcid.org/0000-0003-2361-5510}

Erika Palmerio \orcid{https://orcid.org/0000-0001-6590-3479}

Marco Pinto \orcid{https://orcid.org/0000-0002-5712-9396}

Nina Dresing \orcid{https://orcid.org/0000-0003-3903-4649}

Christina Cohen \orcid{https://orcid.org/0000-0002-0978-8127}

Raúl Gómez-Herrero \orcid{https://orcid.org/0000-0002-5705-9236}

Jan Gieseler \orcid{https://orcid.org/0000-0003-1848-7067}

Francisco Espinosa Lara \orcid{https://orcid.org/0000-0001-9039-8822}

Ignacio Cernuda \orcid{https://orcid.org/0000-0001-8432-5379}

Claire Vallat \orcid{http://orcid.org/0000-0001-6921-868X}

Olivier Witasse \orcid{https://orcid.org/0000-0003-3461-5604}

Nicolas Altobelli \orcid{https://orcid.org/0000-0003-4244-6302}

\end{flushleft}


\begin{thebibliography}{71}
\expandafter\ifx\csname natexlab\endcsname\relax\def\natexlab#1{#1}\fi

\bibitem[{{Acu{\~n}a} {et~al.}(2008){Acu{\~n}a}, {Curtis}, {Scheifele},
  {Russell}, {Schroeder}, {Szabo}, \& {Luhmann}}]{Acuna2008}
{Acu{\~n}a}, M.~H., {Curtis}, D., {Scheifele}, J.~L., {et~al.} 2008, \ssr, 136,
  203

\bibitem[{Allison {et~al.}(2016)Allison, Amako, Apostolakis, Arce, Asai, Aso,
  Bagli, Bagulya, Banerjee, Barrand, Beck, Bogdanov, Brandt, Brown, Burkhardt,
  Canal, Cano-Ott, Chauvie, Cho, Cirrone, Cooperman, Cortés-Giraldo, Cosmo,
  Cuttone, Depaola, Desorgher, Dong, Dotti, Elvira, Folger, Francis, Galoyan,
  Garnier, Gayer, Genser, Grichine, Guatelli, Guèye, Gumplinger, Howard,
  Hřivnáčová, Hwang, Incerti, Ivanchenko, Ivanchenko, Jones, Jun,
  Kaitaniemi, Karakatsanis, Karamitros, Kelsey, Kimura, Koi, Kurashige,
  Lechner, Lee, Longo, Maire, Mancusi, Mantero, Mendoza, Morgan, Murakami,
  Nikitina, Pandola, Paprocki, Perl, Petrović, Pia, Pokorski, Quesada, Raine,
  Reis, Ribon, {Ristić Fira}, Romano, Russo, Santin, Sasaki, Sawkey, Shin,
  Strakovsky, Taborda, Tanaka, Tomé, Toshito, Tran, Truscott, Urban, Uzhinsky,
  Verbeke, Verderi, Wendt, Wenzel, Wright, Wright, Yamashita, Yarba, \&
  Yoshida}]{Geant4Allison}
Allison, J., Amako, K., Apostolakis, J., {et~al.} 2016, Nuclear Instruments and
  Methods in Physics Research Section A: Accelerators, Spectrometers, Detectors
  and Associated Equipment, 835, 186

\bibitem[{{Benkhoff} {et~al.}(2021){Benkhoff}, {Murakami}, {Baumjohann},
  {Besse}, {Bunce}, {Casale}, {Cremonese}, {Glassmeier}, {Hayakawa}, {Heyner},
  {Hiesinger}, {Huovelin}, {Hussmann}, {Iafolla}, {Iess}, {Kasaba},
  {Kobayashi}, {Milillo}, {Mitrofanov}, {Montagnon}, {Novara}, {Orsini},
  {Quemerais}, {Reininghaus}, {Saito}, {Santoli}, {Stramaccioni}, {Sutherland},
  {Thomas}, {Yoshikawa}, \& {Zender}}]{Benkhoff2021}
{Benkhoff}, J., {Murakami}, G., {Baumjohann}, W., {et~al.} 2021, \ssr, 217, 90

\bibitem[{Boggs \& Rogers(1990)}]{Boggs1990}
Boggs, P.~T. \& Rogers, J.~E. 1990, in Contemp. Math., Vol. 112, Statistical
  analysis of measurement error models and applications ({A}rcata, {CA}, 1989)
  (Amer. Math. Soc., Providence, RI), 183--194

\bibitem[{{Burlaga} {et~al.}(1981){Burlaga}, {Sittler}, {Mariani}, \&
  {Schwenn}}]{Burlaga1981}
{Burlaga}, L., {Sittler}, E., {Mariani}, F., \& {Schwenn}, R. 1981, \jgr, 86,
  6673

\bibitem[{{Case} {et~al.}(2020){Case}, {Kasper}, {Stevens}, {Korreck},
  {Paulson}, {Daigneau}, {Caldwell}, {Freeman}, {Henry}, {Klingensmith},
  {Bookbinder}, {Robinson}, {Berg}, {Tiu}, {Wright}, {Reinhart}, {Curtis},
  {Ludlam}, {Larson}, {Whittlesey}, {Livi}, {Klein}, \&
  {Martinovi{\'c}}}]{Case2020}
{Case}, A.~W., {Kasper}, J.~C., {Stevens}, M.~L., {et~al.} 2020, \apjs, 246, 43

\bibitem[{{Domingo} {et~al.}(1995){Domingo}, {Fleck}, \&
  {Poland}}]{Domingo1995}
{Domingo}, V., {Fleck}, B., \& {Poland}, A.~I. 1995, \solphys, 162, 1

\bibitem[{{Dresing} {et~al.}(2020){Dresing}, {Effenberger},
  {G{\'o}mez-Herrero}, {Heber}, {Klassen}, {Kollhoff}, {Richardson}, \&
  {Theesen}}]{Dresing2020}
{Dresing}, N., {Effenberger}, F., {G{\'o}mez-Herrero}, R., {et~al.} 2020, \apj,
  889, 143

\bibitem[{{Dresing} {et~al.}(2014){Dresing}, {G{\'o}mez-Herrero}, {Heber},
  {Klassen}, {Malandraki}, {Dr{\"o}ge}, \& {Kartavykh}}]{Dresing2014}
{Dresing}, N., {G{\'o}mez-Herrero}, R., {Heber}, B., {et~al.} 2014, \aap, 567,
  A27

\bibitem[{{Dresing} {et~al.}(2023){Dresing}, {Rodr{\'\i}guez-Garc{\'\i}a},
  {Jebaraj}, {Warmuth}, {Wallace}, {Balmaceda}, {Podladchikova}, {Strauss},
  {Kouloumvakos}, {Palmroos}, {Krupar}, {Gieseler}, {Xu}, {Mitchell}, {Cohen},
  {de Nolfo}, {Palmerio}, {Carcaboso}, {Kilpua}, {Trotta}, {Auster},
  {Asvestari}, {da Silva}, {Dr{\"o}ge}, {Getachew}, {G{\'o}mez-Herrero},
  {Grande}, {Heyner}, {Holmstr{\"o}m}, {Huovelin}, {Kartavykh}, {Laurenza},
  {Lee}, {Mason}, {Maksimovic}, {Mieth}, {Murakami}, {Oleynik}, {Pinto},
  {Pulupa}, {Richter}, {Rodr{\'\i}guez-Pacheco}, {S{\'a}nchez-Cano},
  {Schuller}, {Ueno}, {Vainio}, {Vecchio}, {Veronig}, \&
  {Wijsen}}]{Dresing2023}
{Dresing}, N., {Rodr{\'\i}guez-Garc{\'\i}a}, L., {Jebaraj}, I.~C., {et~al.}
  2023, \aap, 674, A105

\bibitem[{{Dresing, N.} {et~al.}(2025){Dresing, N.}, {Jebaraj, I. C.}, {Wijsen,
  N.}, {Palmerio, E.}, {Rodríguez-García, L.}, {Palmroos, C.}, {Gieseler,
  J.}, {Jarry, M.}, {Asvestari, E.}, {Mitchell, J. G.}, {Cohen, C. M. S.},
  {Lee, C. O.}, {Wei, W.}, {Ramstad, R.}, {Riihonen, E.}, {Oleynik, P.},
  {Kouloumvakos, A.}, {Warmuth, A.}, {Sánchez-Cano, B.}, {Ehresmann, B.},
  {Dunn, P.}, {Dudnik, O.}, \& {Mac Cormack, C.}}]{Dresing2025}
{Dresing, N.}, {Jebaraj, I. C.}, {Wijsen, N.}, {et~al.} 2025, A\&A, 695, A127

\bibitem[{{Dumbovi{\'c}} {et~al.}(2019){Dumbovi{\'c}}, {Guo}, {Temmer}, {Mays},
  {Veronig}, {Heinemann}, {Dissauer}, {Hofmeister}, {Halekas}, {M{\"o}stl},
  {Amerstorfer}, {Hinterreiter}, {Banjac}, {Herbst}, {Wang}, {Holzknecht},
  {Leitner}, \& {Wimmer{\textendash}Schweingruber}}]{Dumbovic2019}
{Dumbovi{\'c}}, M., {Guo}, J., {Temmer}, M., {et~al.} 2019, \apj, 880, 18

\bibitem[{{Fox} {et~al.}(2016){Fox}, {Velli}, {Bale}, {Decker}, {Driesman},
  {Howard}, {Kasper}, {Kinnison}, {Kusterer}, {Lario}, {Lockwood}, {McComas},
  {Raouafi}, \& {Szabo}}]{Fox2016}
{Fox}, N.~J., {Velli}, M.~C., {Bale}, S.~D., {et~al.} 2016, \ssr, 204, 7

\bibitem[{{Galvin} {et~al.}(2008){Galvin}, {Kistler}, {Popecki}, {Farrugia},
  {Simunac}, {Ellis}, {M{\"o}bius}, {Lee}, {Boehm}, {Carroll}, {Crawshaw},
  {Conti}, {Demaine}, {Ellis}, {Gaidos}, {Googins}, {Granoff}, {Gustafson},
  {Heirtzler}, {King}, {Knauss}, {Levasseur}, {Longworth}, {Singer}, {Turco},
  {Vachon}, {Vosbury}, {Widholm}, {Blush}, {Karrer}, {Bochsler}, {Daoudi},
  {Etter}, {Fischer}, {Jost}, {Opitz}, {Sigrist}, {Wurz}, {Klecker}, {Ertl},
  {Seidenschwang}, {Wimmer-Schweingruber}, {Koeten}, {Thompson}, \&
  {Steinfeld}}]{Galvin2008}
{Galvin}, A.~B., {Kistler}, L.~M., {Popecki}, M.~A., {et~al.} 2008, \ssr, 136,
  437

\bibitem[{{Gieseler} {et~al.}(2023){Gieseler}, {Dresing}, {Palmroos}, {Freiherr
  von Forstner}, {Price}, {Vainio}, {Kouloumvakos},
  {Rodr{\'\i}guez-Garc{\'\i}a}, {Trotta}, {G{\'e}not}, {Masson}, {Roth}, \&
  {Veronig}}]{Gieseler2023}
{Gieseler}, J., {Dresing}, N., {Palmroos}, C., {et~al.} 2023, \frass, 9, 384

\bibitem[{{Gold} {et~al.}(1998){Gold}, {Krimigis}, {Hawkins}, {Haggerty},
  {Lohr}, {Fiore}, {Armstrong}, {Holland}, \& {Lanzerotti}}]{Gold1998}
{Gold}, R.~E., {Krimigis}, S.~M., {Hawkins}, S.~E., I., {et~al.} 1998, \ssr,
  86, 541

\bibitem[{{Grasset} {et~al.}(2013){Grasset}, {Dougherty}, {Coustenis}, {Bunce},
  {Erd}, {Titov}, {Blanc}, {Coates}, {Drossart}, {Fletcher}, {Hussmann},
  {Jaumann}, {Krupp}, {Lebreton}, {Prieto-Ballesteros}, {Tortora}, {Tosi}, \&
  {Van Hoolst}}]{Grasset2013}
{Grasset}, O., {Dougherty}, M.~K., {Coustenis}, A., {et~al.} 2013, \planss, 78,
  1

\bibitem[{Hajdas {et~al.}(2025)Hajdas, Gon{\c c}alves, Pinto, Socha,
  Marcinkowski, Xiao, Santos, Arruda, Galli, Marques, Stein, Meier,
  S{\'a}nchez-Cano, Roussos, Santin, Verstaen, Witasse, Nieminen, Altobelli,
  Desorgher, Menicucci, Reinaecker, Vogiatzi, Rostomyan, \&
  Mrigatshi}]{RADEM2025}
Hajdas, W., Gon{\c c}alves, P., Pinto, M., {et~al.} 2025, Space Science
  Reviews, 221, 43

\bibitem[{{Hajra} {et~al.}(2024){Hajra}, {Tsurutani}, {Lakhina}, {Lu}, \&
  {Du}}]{Hajra2024}
{Hajra}, R., {Tsurutani}, B.~T., {Lakhina}, G.~S., {Lu}, Q., \& {Du}, A. 2024,
  \apj, 974, 264

\bibitem[{{Hapgood}(1992)}]{Hapgood1992}
{Hapgood}, M.~A. 1992, \planss, 40, 711

\bibitem[{{Hayakawa} {et~al.}(2025){Hayakawa}, {Ebihara}, {Mishev},
  {Koldobskiy}, {Kusano}, {Bechet}, {Yashiro}, {Iwai}, {Shinbori}, {Mursula},
  {Miyake}, {Shiota}, {Silveira}, {Stuart}, {Oliveira}, {Akiyama}, {Ohnishi},
  {Ledvina}, \& {Miyoshi}}]{Hayakawa2025}
{Hayakawa}, H., {Ebihara}, Y., {Mishev}, A., {et~al.} 2025, \apj, 979, 49

\bibitem[{{Howard} {et~al.}(2008){Howard}, {Moses}, {Vourlidas}, {Newmark},
  {Socker}, {Plunkett}, {Korendyke}, {Cook}, {Hurley}, {Davila}, {Thompson},
  {St Cyr}, {Mentzell}, {Mehalick}, {Lemen}, {Wuelser}, {Duncan}, {Tarbell},
  {Wolfson}, {Moore}, {Harrison}, {Waltham}, {Lang}, {Davis}, {Eyles},
  {Mapson-Menard}, {Simnett}, {Halain}, {Defise}, {Mazy}, {Rochus}, {Mercier},
  {Ravet}, {Delmotte}, {Auchere}, {Delaboudiniere}, {Bothmer}, {Deutsch},
  {Wang}, {Rich}, {Cooper}, {Stephens}, {Maahs}, {Baugh}, {McMullin}, \&
  {Carter}}]{Howard2008}
{Howard}, R.~A., {Moses}, J.~D., {Vourlidas}, A., {et~al.} 2008, \ssr, 136, 67

\bibitem[{{Jiggens} {et~al.}(2014){Jiggens}, {Chavy-Macdonald}, {Santin},
  {Menicucci}, {Evans}, \& {Hilgers}}]{Jiggens2014}
{Jiggens}, P., {Chavy-Macdonald}, M.-A., {Santin}, G., {et~al.} 2014, \jswsc,
  4, A20

\bibitem[{{Kaiser} {et~al.}(2008){Kaiser}, {Kucera}, {Davila}, {St. Cyr},
  {Guhathakurta}, \& {Christian}}]{Kaiser2008}
{Kaiser}, M.~L., {Kucera}, T.~A., {Davila}, J.~M., {et~al.} 2008, \ssr, 136, 5

\bibitem[{{Kasper} {et~al.}(2016){Kasper}, {Abiad}, {Austin}, {Balat-Pichelin},
  {Bale}, {Belcher}, {Berg}, {Bergner}, {Berthomier}, {Bookbinder}, {Brodu},
  {Caldwell}, {Case}, {Chandran}, {Cheimets}, {Cirtain}, {Cranmer}, {Curtis},
  {Daigneau}, {Dalton}, {Dasgupta}, {DeTomaso}, {Diaz-Aguado}, {Djordjevic},
  {Donaskowski}, {Effinger}, {Florinski}, {Fox}, {Freeman}, {Gallagher},
  {Gary}, {Gauron}, {Gates}, {Goldstein}, {Golub}, {Gordon}, {Gurnee}, {Guth},
  {Halekas}, {Hatch}, {Heerikuisen}, {Ho}, {Hu}, {Johnson}, {Jordan},
  {Korreck}, {Larson}, {Lazarus}, {Li}, {Livi}, {Ludlam}, {Maksimovic},
  {McFadden}, {Marchant}, {Maruca}, {McComas}, {Messina}, {Mercer}, {Park},
  {Peddie}, {Pogorelov}, {Reinhart}, {Richardson}, {Robinson}, {Rosen},
  {Skoug}, {Slagle}, {Steinberg}, {Stevens}, {Szabo}, {Taylor}, {Tiu}, {Turin},
  {Velli}, {Webb}, {Whittlesey}, {Wright}, {Wu}, \& {Zank}}]{Kasper2016}
{Kasper}, J.~C., {Abiad}, R., {Austin}, G., {et~al.} 2016, \ssr, 204, 131

\bibitem[{{Khoo} {et~al.}(2024){Khoo}, {S{\'a}nchez-Cano}, {Lee},
  {Rodr{\'\i}guez-Garc{\'\i}a}, {Kouloumvakos}, {Palmerio}, {Carcaboso},
  {Lario}, {Dresing}, {Cohen}, {McComas}, {Lynch}, {Fraschetti}, {Jebaraj},
  {Mitchell}, {Nieves-Chinchilla}, {Krupar}, {Pacheco}, {Giacalone}, {Auster},
  {Benkhoff}, {Bonnin}, {Christian}, {Ehresmann}, {Fedeli}, {Fischer},
  {Heyner}, {Holmstr{\"o}m}, {Leske}, {Maksimovic}, {Mieth}, {Oleynik},
  {Pinto}, {Richter}, {Rodr{\'\i}guez-Pacheco}, {Schwadron}, {Schmid},
  {Telloni}, {Vecchio}, \& {Wiedenbeck}}]{Khoo2024}
{Khoo}, L.~Y., {S{\'a}nchez-Cano}, B., {Lee}, C.~O., {et~al.} 2024, \apj, 963,
  107

\bibitem[{{Kollhoff} {et~al.}(2021){Kollhoff}, {Kouloumvakos}, {Lario},
  {Dresing}, {G{\'o}mez-Herrero}, {Rodr{\'\i}guez-Garc{\'\i}a}, {Malandraki},
  {Richardson}, {Posner}, {Klein}, {Pacheco}, {Klassen}, {Heber}, {Cohen},
  {Laitinen}, {Cernuda}, {Dalla}, {Espinosa Lara}, {Vainio}, {K{\"o}berle},
  {K{\"u}hl}, {Xu}, {Berger}, {Eldrum}, {Br{\"u}dern}, {Laurenza}, {Kilpua},
  {Aran}, {Rouillard}, {Bu{\v{c}}{\'\i}k}, {Wijsen}, {Pomoell},
  {Wimmer-Schweingruber}, {Martin}, {B{\"o}ttcher}, {Freiherr von Forstner},
  {Terasa}, {Boden}, {Kulkarni}, {Ravanbakhsh}, {Yedla}, {Janitzek},
  {Rodr{\'\i}guez-Pacheco}, {Prieto Mateo}, {S{\'a}nchez Prieto}, {Parra
  Espada}, {Rodr{\'\i}guez Polo}, {Mart{\'\i}nez Hell{\'\i}n}, {Carcaboso},
  {Mason}, {Ho}, {Allen}, {Bruce Andrews}, {Schlemm}, {Seifert}, {Tyagi},
  {Lees}, {Hayes}, {Bale}, {Krupar}, {Horbury}, {Angelini}, {Evans}, {O'Brien},
  {Maksimovic}, {Khotyaintsev}, {Vecchio}, {Steinvall}, \&
  {Asvestari}}]{Kollhoff2021}
{Kollhoff}, A., {Kouloumvakos}, A., {Lario}, D., {et~al.} 2021, \aap, 656, A20

\bibitem[{{Kruparova} {et~al.}(2024){Kruparova}, {Krupar}, {Szabo}, {Lario},
  {Nieves-Chinchilla}, \& {Martinez Oliveros}}]{Kruparova2024}
{Kruparova}, O., {Krupar}, V., {Szabo}, A., {et~al.} 2024, \apjl, 970, L13

\bibitem[{{Lario}(2010)}]{2010Lario}
{Lario}, D. 2010, in American Institute of Physics Conference Series, Vol.
  1216, Twelfth International Solar Wind Conference, ed. M.~{Maksimovic},
  K.~{Issautier}, N.~{Meyer-Vernet}, M.~{Moncuquet}, \& F.~{Pantellini},
  625--628

\bibitem[{{Lario} {et~al.}(2006){Lario}, {Kallenrode}, {Decker}, {Roelof},
  {Krimigis}, {Aran}, \& {Sanahuja}}]{Lario2006}
{Lario}, D., {Kallenrode}, M.-B., {Decker}, R.~B., {et~al.} 2006, \apj, 653,
  1531

\bibitem[{{Lario} {et~al.}(2022){Lario}, {Wijsen}, {Kwon}, {S{\'a}nchez-Cano},
  {Richardson}, {Pacheco}, {Palmerio}, {Stevens}, {Szabo}, {Heyner}, {Dresing},
  {G{\'o}mez-Herrero}, {Carcaboso}, {Aran}, {Afanasiev}, {Vainio}, {Riihonen},
  {Poedts}, {Br{\"u}den}, {Xu}, \& {Kollhoff}}]{Lario2022}
{Lario}, D., {Wijsen}, N., {Kwon}, R.~Y., {et~al.} 2022, \apj, 934, 55

\bibitem[{{Lemen} {et~al.}(2012){Lemen}, {Title}, {Akin}, {Boerner}, {Chou},
  {Drake}, {Duncan}, {Edwards}, {Friedlaender}, {Heyman}, {Hurlburt}, {Katz},
  {Kushner}, {Levay}, {Lindgren}, {Mathur}, {McFeaters}, {Mitchell}, {Rehse},
  {Schrijver}, {Springer}, {Stern}, {Tarbell}, {Wuelser}, {Wolfson}, {Yanari},
  {Bookbinder}, {Cheimets}, {Caldwell}, {Deluca}, {Gates}, {Golub}, {Park},
  {Podgorski}, {Bush}, {Scherrer}, {Gummin}, {Smith}, {Auker}, {Jerram},
  {Pool}, {Soufli}, {Windt}, {Beardsley}, {Clapp}, {Lang}, \&
  {Waltham}}]{Lemen2012}
{Lemen}, J.~R., {Title}, A.~M., {Akin}, D.~J., {et~al.} 2012, \solphys, 275, 17

\bibitem[{{Lepping} {et~al.}(1995){Lepping}, {Ac{\~{u}}na}, {Burlaga},
  {Farrell}, {Slavin}, {Schatten}, {Mariani}, {Ness}, {Neubauer}, {Whang},
  {Byrnes}, {Kennon}, {Panetta}, {Scheifele}, \& {Worley}}]{Lepping1995}
{Lepping}, R.~P., {Ac{\~{u}}na}, M.~H., {Burlaga}, L.~F., {et~al.} 1995, \ssr,
  71, 207

\bibitem[{{Lin} {et~al.}(1995){Lin}, {Anderson}, {Ashford}, {Carlson},
  {Curtis}, {Ergun}, {Larson}, {McFadden}, {McCarthy}, {Parks}, {R{\`e}me},
  {Bosqued}, {Coutelier}, {Cotin}, {D'Uston}, {Wenzel}, {Sanderson}, {Henrion},
  {Ronnet}, \& {Paschmann}}]{Lin1995}
{Lin}, R.~P., {Anderson}, K.~A., {Ashford}, S., {et~al.} 1995, \ssr, 71, 125

\bibitem[{{Liu} {et~al.}(2024){Liu}, {Hu}, {Zhao}, {Chen}, \& {Wang}}]{Liu2024}
{Liu}, Y.~D., {Hu}, H., {Zhao}, X., {Chen}, C., \& {Wang}, R. 2024, \apjl, 974,
  L8

\bibitem[{{Lugaz} {et~al.}(2017){Lugaz}, {Temmer}, {Wang}, \&
  {Farrugia}}]{Lugaz2017}
{Lugaz}, N., {Temmer}, M., {Wang}, Y., \& {Farrugia}, C.~J. 2017, \solphys,
  292, 64

\bibitem[{{Luhmann} {et~al.}(2008){Luhmann}, {Curtis}, {Schroeder}, {McCauley},
  {Lin}, {Larson}, {Bale}, {Sauvaud}, {Aoustin}, {Mewaldt}, {Cummings},
  {Stone}, {Davis}, {Cook}, {Kecman}, {Wiedenbeck}, {von Rosenvinge}, {Acuna},
  {Reichenthal}, {Shuman}, {Wortman}, {Reames}, {Mueller-Mellin}, {Kunow},
  {Mason}, {Walpole}, {Korth}, {Sanderson}, {Russell}, \&
  {Gosling}}]{Luhmann2008}
{Luhmann}, J.~G., {Curtis}, D.~W., {Schroeder}, P., {et~al.} 2008, \ssr, 136,
  117

\bibitem[{{McComas} {et~al.}(2016){McComas}, {Alexander}, {Angold}, {Bale},
  {Beebe}, {Birdwell}, {Boyle}, {Burgum}, {Burnham}, {Christian}, {Cook},
  {Cooper}, {Cummings}, {Davis}, {Desai}, {Dickinson}, {Dirks}, {Do}, {Fox},
  {Giacalone}, {Gold}, {Gurnee}, {Hayes}, {Hill}, {Kasper}, {Kecman}, {Klemic},
  {Krimigis}, {Labrador}, {Layman}, {Leske}, {Livi}, {Matthaeus}, {McNutt},
  {Mewaldt}, {Mitchell}, {Nelson}, {Parker}, {Rankin}, {Roelof}, {Schwadron},
  {Seifert}, {Shuman}, {Stokes}, {Stone}, {Vandegriff}, {Velli}, {von
  Rosenvinge}, {Weidner}, {Wiedenbeck}, \& {Wilson}}]{McComas2016}
{McComas}, D.~J., {Alexander}, N., {Angold}, N., {et~al.} 2016, \ssr, 204, 187

\bibitem[{{McKibben}(1972)}]{1972McKibben}
{McKibben}, R.~B. 1972, \jgr, 77, 3957

\bibitem[{{Mewaldt} {et~al.}(2008){Mewaldt}, {Cohen}, {Cook}, {Cummings},
  {Davis}, {Geier}, {Kecman}, {Klemic}, {Labrador}, {Leske}, {Miyasaka},
  {Nguyen}, {Ogliore}, {Stone}, {Radocinski}, {Wiedenbeck}, {Hawk}, {Shuman},
  {von Rosenvinge}, \& {Wortman}}]{Mewaldt2008}
{Mewaldt}, R.~A., {Cohen}, C.~M.~S., {Cook}, W.~R., {et~al.} 2008, \ssr, 136,
  285

\bibitem[{{M{\"u}ller} {et~al.}(2020){M{\"u}ller}, {St. Cyr}, {Zouganelis},
  {Gilbert}, {Marsden}, {Nieves-Chinchilla}, {Antonucci}, {Auch{\`e}re},
  {Berghmans}, {Horbury}, {Howard}, {Krucker}, {Maksimovic}, {Owen}, {Rochus},
  {Rodriguez-Pacheco}, {Romoli}, {Solanki}, {Bruno}, {Carlsson}, {Fludra},
  {Harra}, {Hassler}, {Livi}, {Louarn}, {Peter}, {Sch{\"u}hle}, {Teriaca}, {del
  Toro Iniesta}, {Wimmer-Schweingruber}, {Marsch}, {Velli}, {De Groof},
  {Walsh}, \& {Williams}}]{Muller2020}
{M{\"u}ller}, D., {St. Cyr}, O.~C., {Zouganelis}, I., {et~al.} 2020, \aap, 642,
  A1

\bibitem[{{M{\"u}ller-Mellin} {et~al.}(1995){M{\"u}ller-Mellin}, {Kunow},
  {Flei{\ss}ner}, {Pehlke}, {Rode}, {R{\"o}schmann}, {Scharmberg}, {Sierks},
  {Rusznyak}, {Mckenna-Lawlor}, {Elendt}, {Sequeiros}, {Meziat}, {Sanchez},
  {Medina}, {del Peral}, {Witte}, {Marsden}, \& {Henrion}}]{Muller-Mellin1995}
{M{\"u}ller-Mellin}, R., {Kunow}, H., {Flei{\ss}ner}, V., {et~al.} 1995,
  \solphys, 162, 483

\bibitem[{{Odstrcil} {et~al.}(2004){Odstrcil}, {Riley}, \&
  {Zhao}}]{Odstrcil2004}
{Odstrcil}, D., {Riley}, P., \& {Zhao}, X.~P. 2004, Journal of Geophysical
  Research (Space Physics), 109, A02116

\bibitem[{{Ogilvie} {et~al.}(1995){Ogilvie}, {Chornay}, {Fritzenreiter},
  {Hunsaker}, {Keller}, {Lobell}, {Miller}, {Scudder}, {Sittler}, {Torbert},
  {Bodet}, {Needell}, {Lazarus}, {Steinberg}, {Tappan}, {Mavretic}, \&
  {Gergin}}]{Ogilvie1995}
{Ogilvie}, K.~W., {Chornay}, D.~J., {Fritzenreiter}, R.~J., {et~al.} 1995,
  \ssr, 71, 55

\bibitem[{{Ogilvie} \& {Desch}(1997)}]{Ogilvie1997}
{Ogilvie}, K.~W. \& {Desch}, M.~D. 1997, \adv, 20, 559

\bibitem[{{Owen} {et~al.}(2020){Owen}, {Bruno}, {Livi}, {Louarn}, {Al Janabi},
  {Allegrini}, {Amoros}, {Baruah}, {Barthe}, {Berthomier}, {Bordon},
  {Brockley-Blatt}, {Brysbaert}, {Capuano}, {Collier}, {DeMarco}, {Fedorov},
  {Ford}, {Fortunato}, {Fratter}, {Galvin}, {Hancock}, {Heirtzler}, {Kataria},
  {Kistler}, {Lepri}, {Lewis}, {Loeffler}, {Marty}, {Mathon}, {Mayall}, {Mele},
  {Ogasawara}, {Orlandi}, {Pacros}, {Penou}, {Persyn}, {Petiot}, {Phillips},
  {P{\v{r}}ech}, {Raines}, {Reden}, {Rouillard}, {Rousseau}, {Rubiella},
  {Seran}, {Spencer}, {Thomas}, {Trevino}, {Verscharen}, {Wurz}, {Alapide},
  {Amoruso}, {Andr{\'e}}, {Anekallu}, {Arciuli}, {Arnett}, {Ascolese},
  {Bancroft}, {Bland}, {Brysch}, {Calvanese}, {Castronuovo},
  {{\v{C}}erm{\'a}k}, {Chornay}, {Clemens}, {Coker}, {Collinson}, {D'Amicis},
  {Dandouras}, {Darnley}, {Davies}, {Davison}, {De Los Santos}, {Devoto},
  {Dirks}, {Edlund}, {Fazakerley}, {Ferris}, {Frost}, {Fruit}, {Garat},
  {G{\'e}not}, {Gibson}, {Gilbert}, {de Giosa}, {Gradone}, {Hailey}, {Horbury},
  {Hunt}, {Jacquey}, {Johnson}, {Lavraud}, {Lawrenson}, {Leblanc}, {Lockhart},
  {Maksimovic}, {Malpus}, {Marcucci}, {Mazelle}, {Monti}, {Myers}, {Nguyen},
  {Rodriguez-Pacheco}, {Phillips}, {Popecki}, {Rees}, {Rogacki}, {Ruane},
  {Rust}, {Salatti}, {Sauvaud}, {Stakhiv}, {Stange}, {Stubbs}, {Taylor},
  {Techer}, {Terrier}, {Thibodeaux}, {Urdiales}, {Varsani}, {Walsh}, {Watson},
  {Wheeler}, {Willis}, {Wimmer-Schweingruber}, {Winter}, {Yardley}, \&
  {Zouganelis}}]{Owen2020}
{Owen}, C.~J., {Bruno}, R., {Livi}, S., {et~al.} 2020, \aap, 642, A16

\bibitem[{{Palmerio} {et~al.}(2021){Palmerio}, {Kilpua}, {Witasse}, {Barnes},
  {S{\'a}nchez-Cano}, {Weiss}, {Nieves-Chinchilla}, {M{\"o}stl}, {Jian},
  {Mierla}, {Zhukov}, {Guo}, {Rodriguez}, {Lowrance}, {Isavnin}, {Turc},
  {Futaana}, \& {Holmstr{\"o}m}}]{Palmerio2021}
{Palmerio}, E., {Kilpua}, E. K.~J., {Witasse}, O., {et~al.} 2021, Space
  Weather, 19, e2020SW002654

\bibitem[{{Pesnell} {et~al.}(2012){Pesnell}, {Thompson}, \&
  {Chamberlin}}]{Pesnell2012}
{Pesnell}, W.~D., {Thompson}, B.~J., \& {Chamberlin}, P.~C. 2012, \solphys,
  275, 3

\bibitem[{Pinto(2019)}]{PintoThesis2019}
Pinto, M. 2019, PhD thesis, University of Lisbon - Instituto Superior Tecnico

\bibitem[{{Pinto} {et~al.}(2020){Pinto}, {Goncalves}, {Hajdas}, \&
  {Socha}}]{Pinto2020}
{Pinto}, M., {Goncalves}, P., {Hajdas}, W., \& {Socha}, P. 2020, in European
  Planetary Science Congress, EPSC2020--311

\bibitem[{{Raukunen} {et~al.}(2020){Raukunen}, {Paassilta}, {Vainio},
  {Rodriguez}, {Eronen}, {Crosby}, {Dierckxsens}, {Jiggens}, {Heynderickx}, \&
  {Sandberg}}]{Raukunen2020}
{Raukunen}, O., {Paassilta}, M., {Vainio}, R., {et~al.} 2020, Journal of Space
  Weather and Space Climate, 10, 24

\bibitem[{{Richardson} \& {Cane}(1996)}]{RichardsonCane1996}
{Richardson}, I.~G. \& {Cane}, H.~V. 1996, \jgr, 101, 27521

\bibitem[{{Rodr{\'\i}guez-Garc{\'\i}a}
  {et~al.}(2025){Rodr{\'\i}guez-Garc{\'\i}a}, {G{\'o}mez-Herrero}, {Dresing},
  {Balmaceda}, {Palmerio}, {Kouloumvakos}, {Jebaraj}, {Espinosa Lara}, {Roco},
  {Palmroos}, {Warmuth}, {Nicolaou}, {Mason}, {Guo}, {Laitinen}, {Cernuda},
  {Nieves-Chinchilla}, {Fedeli}, {Lee}, {Cohen}, {Owen}, {Ho}, {Malandraki},
  {Vainio}, \& {Rodr{\'\i}guez-Pacheco}}]{Rodriguez-Garcia2025}
{Rodr{\'\i}guez-Garc{\'\i}a}, L., {G{\'o}mez-Herrero}, R., {Dresing}, N.,
  {et~al.} 2025, \aap, 694, A64

\bibitem[{{Rodr{\'\i}guez-Garc{\'\i}a}
  {et~al.}(2023){Rodr{\'\i}guez-Garc{\'\i}a}, {G{\'o}mez-Herrero}, {Dresing},
  {Lario}, {Zouganelis}, {Balmaceda}, {Kouloumvakos}, {Fedeli}, {Espinosa
  Lara}, {Cernuda}, {Ho}, {Wimmer-Schweingruber}, \&
  {Rodr{\'\i}guez-Pacheco}}]{Rodriguez-Garcia2023a}
{Rodr{\'\i}guez-Garc{\'\i}a}, L., {G{\'o}mez-Herrero}, R., {Dresing}, N.,
  {et~al.} 2023, \aap, 670, A51

\bibitem[{{Rodr{\'\i}guez-Garc{\'\i}a}
  {et~al.}(2021){Rodr{\'\i}guez-Garc{\'\i}a}, {G{\'o}mez-Herrero},
  {Zouganelis}, {Balmaceda}, {Nieves-Chinchilla}, {Dresing}, {Dumbovi{\'c}},
  {Nitta}, {Carcaboso}, {dos Santos}, {Jian}, {Mays}, {Williams}, \&
  {Rodr{\'\i}guez-Pacheco}}]{Rodriguez-Garcia2021}
{Rodr{\'\i}guez-Garc{\'\i}a}, L., {G{\'o}mez-Herrero}, R., {Zouganelis}, I.,
  {et~al.} 2021, \aap, 653, A137

\bibitem[{{Rodr{\'\i}guez-Pacheco} {et~al.}(2020){Rodr{\'\i}guez-Pacheco},
  {Wimmer-Schweingruber}, {Mason}, {Ho}, {S{\'a}nchez-Prieto}, {Prieto},
  {Mart{\'\i}n}, {Seifert}, {Andrews}, {Kulkarni}, {Panitzsch}, {Boden},
  {B{\"o}ttcher}, {Cernuda}, {Elftmann}, {Espinosa Lara}, {G{\'o}mez-Herrero},
  {Terasa}, {Almena}, {Begley}, {B{\"o}hm}, {Blanco}, {Boogaerts}, {Carrasco},
  {Castillo}, {da Silva Fari{\~n}a}, {de Manuel Gonz{\'a}lez}, {Drews},
  {Dupont}, {Eldrum}, {Gordillo}, {Guti{\'e}rrez}, {Haggerty}, {Hayes},
  {Heber}, {Hill}, {J{\"u}ngling}, {Kerem}, {Knierim}, {K{\"o}hler}, {Kolbe},
  {Kulemzin}, {Lario}, {Lees}, {Liang}, {Mart{\'\i}nez Hell{\'\i}n}, {Meziat},
  {Montalvo}, {Nelson}, {Parra}, {Paspirgilis}, {Ravanbakhsh}, {Richards},
  {Rodr{\'\i}guez-Polo}, {Russu}, {S{\'a}nchez}, {Schlemm}, {Schuster},
  {Seimetz}, {Steinhagen}, {Tammen}, {Tyagi}, {Varela}, {Yedla}, {Yu},
  {Agueda}, {Aran}, {Horbury}, {Klecker}, {Klein}, {Kontar}, {Krucker},
  {Maksimovic}, {Malandraki}, {Owen}, {Pacheco}, {Sanahuja}, {Vainio},
  {Connell}, {Dalla}, {Dr{\"o}ge}, {Gevin}, {Gopalswamy}, {Kartavykh},
  {Kudela}, {Limousin}, {Makela}, {Mann}, {{\"O}nel}, {Posner}, {Ryan},
  {Soucek}, {Hofmeister}, {Vilmer}, {Walsh}, {Wang}, {Wiedenbeck}, {Wirth}, \&
  {Zong}}]{Rodriguez-Pacheco2020}
{Rodr{\'\i}guez-Pacheco}, J., {Wimmer-Schweingruber}, R.~F., {Mason}, G.~M.,
  {et~al.} 2020, \aap, 642, A7

\bibitem[{{Roelof} {et~al.}(1992){Roelof}, {Gold}, {Simnett}, {Tappin},
  {Armstrong}, \& {Lanzerotti}}]{Roelof1992}
{Roelof}, E.~C., {Gold}, R.~E., {Simnett}, G.~M., {et~al.} 1992, \grl, 19, 1243

\bibitem[{{S{\'a}nchez-Cano} {et~al.}(2023){S{\'a}nchez-Cano}, {Witasse},
  {Knutsen}, {Meggi}, {Viet}, {Lester}, {Wimmer-Schweingruber}, {Pinto},
  {Moissl}, {Benkhoff}, {Opgenoorth}, {Auster}, {de Brujine}, {Collins}, {De
  Marchi}, {Fischer}, {Futaana}, {Godfrey}, {Heyner}, {Holmstrom}, {Johnstone},
  {Joyce}, {Lakey}, {Martinez}, {Milligan}, {Montagnon}, {M{\"u}ller}, {Livi},
  {Prusti}, {Raines}, {Richter}, {Schmid}, {Schmitz}, {Svedhem}, {Taylor},
  {Tremolizzo}, {Titov}, {Wilson}, {Wood}, \& {Zender}}]{Sanchez-Cano2023}
{S{\'a}nchez-Cano}, B., {Witasse}, O., {Knutsen}, E.~W., {et~al.} 2023, Space
  Weather, 21, e2023SW003540

\bibitem[{{Sauvaud} {et~al.}(2008){Sauvaud}, {Larson}, {Aoustin}, {Curtis},
  {M{\'e}dale}, {Fedorov}, {Rouzaud}, {Luhmann}, {Moreau}, {Schr{\"o}der},
  {Louarn}, {Dandouras}, \& {Penou}}]{Sauvaud2008}
{Sauvaud}, J.~A., {Larson}, D., {Aoustin}, C., {et~al.} 2008, \ssr, 136, 227

\bibitem[{{Stone} {et~al.}(1998){Stone}, {Frandsen}, {Mewaldt}, {Christian},
  {Margolies}, {Ormes}, \& {Snow}}]{Stone1998}
{Stone}, E.~C., {Frandsen}, A.~M., {Mewaldt}, R.~A., {et~al.} 1998, \ssr, 86, 1

\bibitem[{{Strauss} {et~al.}(2020){Strauss}, {Dresing}, {Kollhoff}, \&
  {Br{\"u}dern}}]{Strauss2020}
{Strauss}, R.~D., {Dresing}, N., {Kollhoff}, A., \& {Br{\"u}dern}, M. 2020,
  \apj, 897, 24

\bibitem[{{Thernisien} {et~al.}(2009){Thernisien}, {Vourlidas}, \&
  {Howard}}]{Thernisien2009}
{Thernisien}, A., {Vourlidas}, A., \& {Howard}, R.~A. 2009, \solphys, 256, 111

\bibitem[{{Thernisien} {et~al.}(2006){Thernisien}, {Howard}, \&
  {Vourlidas}}]{Thernisien2006}
{Thernisien}, A.~F.~R., {Howard}, R.~A., \& {Vourlidas}, A. 2006, \apj, 652,
  763

\bibitem[{{Torsti} {et~al.}(1995){Torsti}, {Valtonen}, {Lumme}, {Peltonen},
  {Eronen}, {Louhola}, {Riihonen}, {Schultz}, {Teittinen}, {Ahola}, {Holmlund},
  {Kelh{\"a}}, {Lepp{\"a}l{\"a}}, {Ruuska}, \& {Str{\"o}mmer}}]{Torsti1995}
{Torsti}, J., {Valtonen}, E., {Lumme}, M., {et~al.} 1995, \solphys, 162, 505

\bibitem[{{Van Allen} {et~al.}(1974){Van Allen}, {Baker}, {Randall}, \&
  {Sentman}}]{1974VanAllen}
{Van Allen}, J.~A., {Baker}, D.~N., {Randall}, B.~A., \& {Sentman}, D.~D. 1974,
  \jgr, 79, 3559

\bibitem[{Virtanen {et~al.}(2020)Virtanen, Gommers, Oliphant, Haberland, Reddy,
  Cournapeau, Burovski, Peterson, Weckesser, Bright, {van der Walt}, Brett,
  Wilson, Millman, Mayorov, Nelson, Jones, Kern, Larson, Carey, Polat, Feng,
  Moore, {VanderPlas}, Laxalde, Perktold, Cimrman, Henriksen, Quintero, Harris,
  Archibald, Ribeiro, Pedregosa, {van Mulbregt}, \& {SciPy 1.0
  Contributors}}]{Virtanen2020}
Virtanen, P., Gommers, R., Oliphant, T.~E., {et~al.} 2020, Nature Methods, 17,
  261

\bibitem[{{von Rosenvinge} {et~al.}(2008){von Rosenvinge}, {Reames}, {Baker},
  {Hawk}, {Nolan}, {Ryan}, {Shuman}, {Wortman}, {Mewaldt}, {Cummings}, {Cook},
  {Labrador}, {Leske}, \& {Wiedenbeck}}]{vonRosenvinge2008}
{von Rosenvinge}, T.~T., {Reames}, D.~V., {Baker}, R., {et~al.} 2008, \ssr,
  136, 391

\bibitem[{{Weiler} {et~al.}(2025){Weiler}, {M{\"o}stl}, {Davies}, {Veronig},
  {Amerstorfer}, {Amerstorfer}, {Le Lou{\"e}dec}, {Bauer}, {Lugaz}, {Haberle},
  {R{\"u}disser}, {Majumdar}, \& {Reiss}}]{Weiler2025}
{Weiler}, E., {M{\"o}stl}, C., {Davies}, E.~E., {et~al.} 2025, Space Weather,
  23, e2024SW004260

\bibitem[{{Whitman} {et~al.}(2023){Whitman}, {Egeland}, {Richardson},
  {Allison}, {Quinn}, {Barzilla}, {Kitiashvili}, {Sadykov}, {Bain},
  {Dierckxsens}, {Mays}, {Tadesse}, {Lee}, {Semones}, {Luhmann},
  {N{\'u}{\~n}ez}, {White}, {Kahler}, {Ling}, {Smart}, {Shea}, {Tenishev},
  {Boubrahimi}, {Aydin}, {Martens}, {Angryk}, {Marsh}, {Dalla}, {Crosby},
  {Schwadron}, {Kozarev}, {Gorby}, {Young}, {Laurenza}, {Cliver}, {Alberti},
  {Stumpo}, {Benella}, {Papaioannou}, {Anastasiadis}, {Sandberg}, {Georgoulis},
  {Ji}, {Kempton}, {Pandey}, {Li}, {Hu}, {Zank}, {Lavasa}, {Giannopoulos},
  {Falconer}, {Kadadi}, {Fernandes}, {Dayeh}, {Mu{\~n}oz-Jaramillo},
  {Chatterjee}, {Moreland}, {Sokolov}, {Roussev}, {Taktakishvili},
  {Effenberger}, {Gombosi}, {Huang}, {Zhao}, {Wijsen}, {Aran}, {Poedts},
  {Kouloumvakos}, {Paassilta}, {Vainio}, {Belov}, {Eroshenko}, {Abunina},
  {Abunin}, {Balch}, {Malandraki}, {Karavolos}, {Heber}, {Labrenz}, {K{\"u}hl},
  {Kosovichev}, {Oria}, {Nita}, {Illarionov}, {O'Keefe}, {Jiang}, {Fereira},
  {Ali}, {Paouris}, {Aminalragia-Giamini}, {Jiggens}, {Jin}, {Lee}, {Palmerio},
  {Bruno}, {Kasapis}, {Wang}, {Chen}, {Sanahuja}, {Lario}, {Jacobs}, {Strauss},
  {Steyn}, {van den Berg}, {Swalwell}, {Waterfall}, {Nedal}, {Miteva},
  {Dechev}, {Zucca}, {Engell}, {Maze}, {Farmer}, {Kerber}, {Barnett}, {Loomis},
  {Grey}, {Thompson}, {Linker}, {Caplan}, {Downs}, {T{\"o}r{\"o}k}, {Lionello},
  {Titov}, {Zhang}, \& {Hosseinzadeh}}]{Whitman2023}
{Whitman}, K., {Egeland}, R., {Richardson}, I.~G., {et~al.} 2023, \adv, 72,
  5161

\bibitem[{{Wiedenbeck} {et~al.}(2017){Wiedenbeck}, {Angold}, {Birdwell},
  {Burnham}, {Christian}, {Cohen}, {Cook}, {Cummings}, {Davis}, {Dirks}, {Do},
  {Everett}, {Goodwin}, {Hanley}, {Hernandez}, {Kecman}, {Klemic}, {Labrador},
  {Leske}, {Lopez}, {Link}, {McComas}, {Mewaldt}, {Miyasaka}, {Nahory},
  {Rankin}, {Riggans}, {Rodriguez}, {Rusert}, {Shuman}, {Simms}, {Stone}, {von
  Rosenvinge}, {Weidner}, \& {White}}]{Wiedenbeck2017}
{Wiedenbeck}, M.~E., {Angold}, N.~G., {Birdwell}, B., {et~al.} 2017, in
  International Cosmic Ray Conference, Vol. 301, 35th International Cosmic Ray
  Conference (ICRC2017), 16

\bibitem[{{Zurbuchen} \& {Richardson}(2006)}]{Zurbuchen2006}
{Zurbuchen}, T.~H. \& {Richardson}, I.~G. 2006, \ssr, 123, 31

\end{thebibliography}

\clearpage
\onecolumn


\begin{appendix}
\section{ENLIL simulation}
\label{app_ENLIL}
ENLIL is a three-dimensional, time-dependent magnetohydrodynamic (MHD) model designed to simulate the heliospheric environment beyond 21.5 solar radii. A detailed description of the model can be found in Appendix A by \cite{Rodriguez-Garcia2025}. The state of the heliosphere and interactions with interplanetary structures present at the time of the SEP event can significantly affect spacecraft magnetic connectivity. To account for these influences, the ENLIL simulation was run from May 8 to May 18, covering five days before and after the event. This period includes earlier CMEs that could affect particle propagation and tracks the evolution of ICMEs through the interplanetary medium out to 2.1 au. The 3D parameters for seven relevant CMEs occurring between May 8 and May 13 were taken from \citet{Liu2024} for use in the simulation. The parameters for the CME and model setup, along with the simulation results, can be accessed on the Community Coordinated Modeling Center (CCMC) website.\footnote{\url{https://ccmc.gsfc.nasa.gov/results/viewrun.php?domain=SH&runnumber=Laura_Rodriguez-Garcia_092424_SH_1}}.

\section{STEREO-A threshold mode}
\label{app:data_gap}

\begin{figure*}[th!]
\centering
  \includegraphics[width=0.8\linewidth]{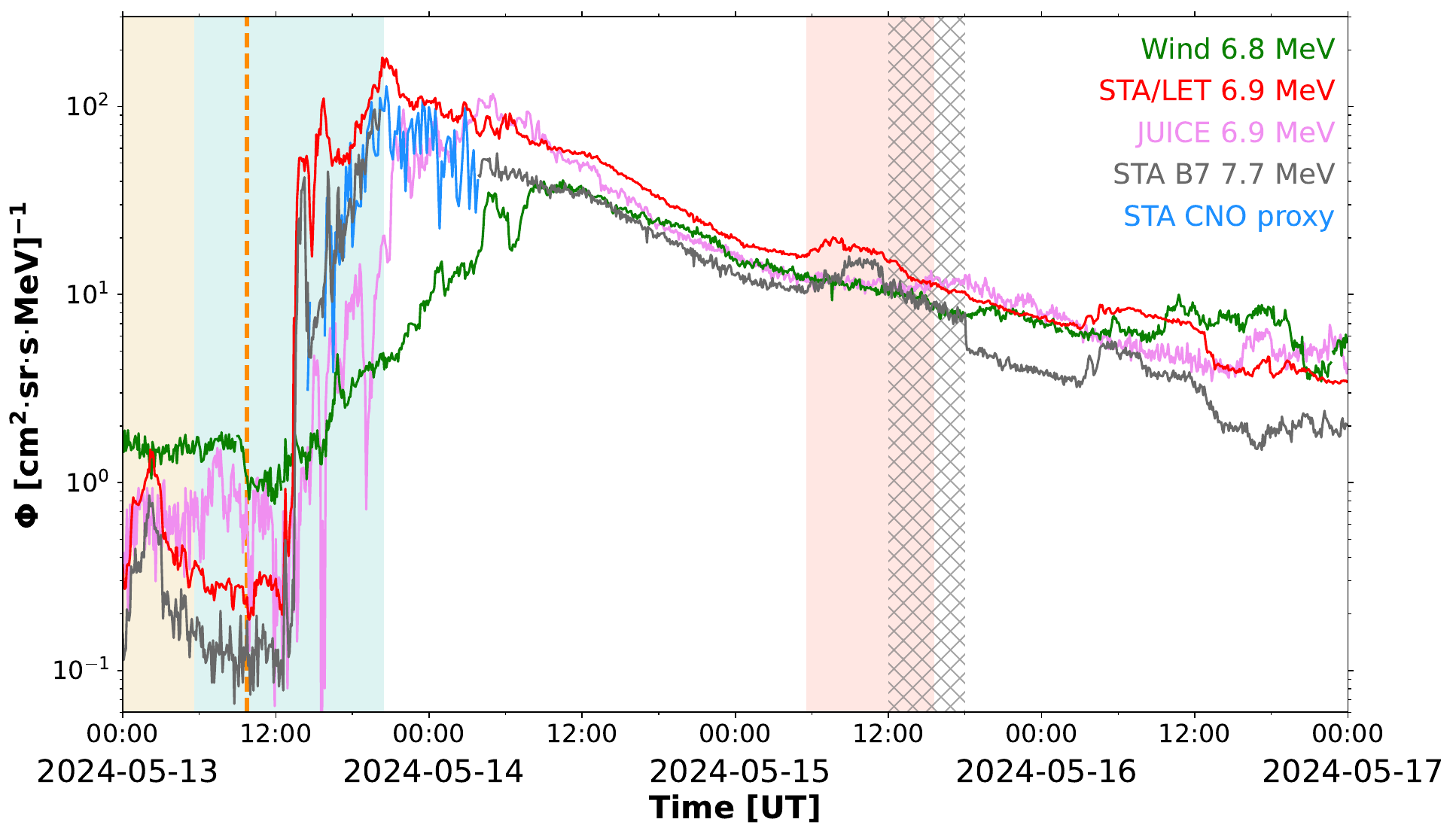}
     \caption{In-situ proton time profiles by Wind/3DP--omni (green), STEREO-A/LET--omni (red), STEREO-A/LET--B7 sector (grey), STEREO-A/LET CNO proxy (blue),  and JUICE/RADEM--anti-Sun (magenta) for ${\sim}6.9$ MeV. Shading and lines are shown as in Fig.~\ref{fig:seps_by_energy}.}
     \label{fig:gap}
\end{figure*}

The 2024 May 13 SEP event was large enough that STEREO-A/LET went into dynamic threshold mode. It means that the B7 sectored data  for protons and He (as that is one of the detectors affected) was not available during mode 2 \citep{Mewaldt2008}. Mode 2 was on from 2024 May 13 at 20:14~UT to May 14 at 03:58~UT, as shown in the data gap in panel (i), grey line, in Fig.~\ref{fig:seps_by_energy}. To estimate the flux measured by STEREO-A/LET/B7 sectored data during the gap, we followed the approximation method explain below. We used the CNO energy sectored rates that were available and took a ratio between the averaged sectors B3 and B4 to sector B7, as CNO should not be affected by the dynamic thresholds. Then, we used this ratio to scale the H averaged sectors B3 and B4 data to get an approximation of what sector B7 should be measuring during this time.  We note that we assumed that CNO has the same anisotropy that H has and that the composition did not change during this time period.   
Figure~\ref{fig:gap} presents the result of this assumption, showing that the CNO proxy follows the trend of the gray line representing the B7 sector. 


\end{appendix}

\end{document}